\documentclass{aa}

\usepackage{graphicx}
\usepackage{txfonts}
\usepackage{bm}
\usepackage{hyperref}
\usepackage{multirow}
\usepackage{textcomp}
\usepackage{xcolor}

\begin{document}

   \title{Mid-infrared photometry of the T Tauri triple system with kernel phase interferometry\thanks{Based on observations made with ESO telescopes at Paranal Observatory under program ID 60.A-9107(E).}}

   \author{J. Kammerer
          \inst{1,2}
          \and
          M. Kasper\inst{1}
          \and
          M. J. Ireland\inst{2}
          \and
          R. K\"ohler\inst{3}
          \and
          R. Laugier\inst{4}
          \and
          F. Martinache\inst{4}
          \and
          R. Siebenmorgen\inst{1}
          \and
          M. E. van den Ancker\inst{1}
          \and
          R. van Boekel\inst{5}
          \and
          T. M. Herbst\inst{5}
          \and
          E. Pantin\inst{6}
          \and
          H.-U. K\"aufl\inst{1}
          \and
          D. J. M. Petit dit de la Roche\inst{1}
          \and
          V. D. Ivanov\inst{1}
          }

   \institute{European Southern Observatory, Karl-Schwarzschild-Str. 2, 85748, Garching, Germany\\
              \email{jens.kammerer@eso.org}
         \and
             Research School of Astronomy \& Astrophysics, Australian National University, Canberra, ACT 2611, Australia
         \and
             University of Vienna, Department of Astrophysics, T\"urkenschanzstr. 17 (Sternwarte), 1180 Vienna, Austria
         \and
             Universit\'e C\^ote d’Azur, Observatoire de la C\^ote d’Azur, CNRS, Laboratoire Lagrange, Lagrange, France
         \and
             Max Planck-Institut f\"ur Astronomie, K\"onigstuhl 17, 69117 Heidelberg, Germany
         \and
             Laboratoire CEA, IRFU/DAp, AIM, Universit\'e Paris-Saclay, Universit\'e Paris Diderot, Sorbonne Paris Cit\'e, CNRS, 91191 Gif-sur-Yvette, France
             }

   \date{Received today; accepted tomorrow}

 
  \abstract
   {T~Tauri has long been the prototypical young pre-main-sequence star. However, with increasing resolution and sensitivity, T~Tauri has now been decomposed into a triple system with a complex disk and outflow geometry.}
   {We aim to measure the brightness of all three components of the T Tauri system (T Tau N, T Tau Sa, and T Tau Sb) in the mid-infrared in order to obtain photometry around the $\sim 9.7~\text{\textmu m}$ silicate feature. This allows us to study their variability and to investigate the distribution of dust and the geometry of circumstellar and circumbinary disks in this complex system.}
   {We observe the T~Tauri system with the Very Large Telescope (VLT)/VISIR-NEAR instrument, performing diffraction-limited imaging in the mid-infrared. With kernel phase interferometry post-processing of the data, and using the astrometric positions of all three components from VLT/SPHERE, we measure the three components' individual brightnesses (including the southern binary at an angular separation down to $\sim 0.2~\lambda/D$) and obtain their photometry. In order to validate our methods, we simulate and recover mock data of the T~Tauri system using the observed reference point-spread function of HD~27639.}
   {We find that T Tau N is rather stable and shows weak silicate emission, while T Tau Sa is highly variable and shows prominent silicate absorption. T Tau Sb became significantly fainter compared to data from 2004 and 2006, suggesting increased extinction by dust. The precision of our photometry is limited by systematic errors in kernel phase interferometry, which is consistent with previous studies using this technique.}
   {Our results confirm the complex scenario of misaligned disks in the T Tauri system that had been observed previously, and they are in agreement with the recently observed dimming of T Tau Sb in the near-infrared. Our mid-infrared photometry supports the interpretation that T Tau Sb has moved behind the dense region of the Sa-Sb circumbinary disk on its tight orbit around Sa, therefore suffering increased extinction.}

   \keywords{Stars: variables: T Tauri --
                (Stars:) binaries (including multiple): close --
                Techniques: high angular resolution
               }

   \maketitle
%
\section{Introduction}
\label{sec:introduction}

T Tauri (T Tau) is the historical prototype of a young and accreting low-mass star. As such, it has provided astronomers with many surprise discoveries over the past decades. While T Tau was initially believed to be a single star surrounded by a circumstellar disk (CSD), from which it accretes matter, \citet{dyck1982} used near-infrared speckle interferometry to show that it is in fact a binary composed of an optically bright northern component (T Tau N) and an optically faint southern component (T Tau S). It was suspected that the optical faintness of T Tau S is caused by extinction from circumstellar material that hides the southern component behind gas and dust. Indeed, \citet{ghez1991} found T Tau S to be variable at all near-infrared to mid-infrared wavelengths and concluded that its spectral energy distribution (SED) must be dominated by variable accretion from a CSD. Five years later, the detection of perpendicular jets expelled from T Tau N and S supported the theory that T Tau is composed of not one, but two young and accreting stars \citep{herbst1996}.

When \citet{koresko2000} observed T Tau S with the Keck Near-IR Camera (NIRC) using speckle interferometry, they found that it is itself a tight binary consisting of T Tau Sa and T Tau Sb, and the detection of significant orbital motion of T Tau Sa around Sb by \citet{duchene2002} implied a high mass ratio between the two southern components. Their medium-resolution spectroscopy further revealed that T Tau Sa has a featureless spectrum and that Sb is an embedded M-type classical T Tauri star (CTTS), suggesting that both southern components are surrounded by dense material. Later, \citet{kasper2002} and \citet{duchene2005} found the CSD of T Tau Sa to be oriented edge-on and attributed the strong optical extinction of $\sim 15~\text{mag}$ around the southern components to circumbinary material (e.g., an Sa-Sb circumbinary disk). \citet{duchene2006} identified T Tau Sa to be the most massive component of the triple system, most likely a young Herbig Ae star, from orbital monitoring.

Around 2005, the triple system was first resolved in the N-band by two teams. \citet{ratzka2009} used the Very Large Telescope Interferometer (VLTI) MID-infrared Interferometric instrument (MIDI) to observe T Tau in November 2004. They found that the silicate band at $9.7~\text{\textmu m}$ is seen in absorption around both southern components, which confirmed a high extinction by Sa-Sb circumbinary material. Moreover, they resolved a small edge-on disk around T Tau Sa, roughly oriented from north to south. \citet{skemer2008} used the Multiple Mirror Telescope (MMT) deformable secondary mirror for adaptive optics-assisted imaging in the mid-infrared in November 2006. The high Strehl ratio and the good point-spread function (PSF) stability allowed them to resolve the southern binary at one-third of the classical diffraction limit of a telescope. Their data allowed them to conclude that the material producing the strong silicate absorption toward T Tau S is entirely in front of Sa. Later, \citet{vanboekel2010} measured a very rapid flux increase at $12.8~\text{\textmu m}$ of the unresolved T Tau S binary over just four days, which they attributed to an increased accretion rate. Slower variations could still be well produced by variable extinction, and the most likely scenario is a combination of both processes. Also in 2005, \citet{monnier2009} observed T Tau in the mid-infrared using a technique called segment-tilting, which is a hybrid between aperture masking and kernel phase interferometry, utilizing almost the entire Keck primary mirror while separating the light into multiple non-redundant interferograms. However, their observations were limited by calibration issues caused by bandwidth smearing and the close southern binary could not be properly resolved.

Through orbital monitoring with the Very Large Telescope (VLT) Spectro-Polarimetric High-contrast Exoplanet Research instrument (SPHERE), \citet{koehler2016} were able to precisely measure the orbits of all three components of the T Tau system, as well as their masses. They found the masses of T Tau Sa and Sb to be $\sim 2.12~\rm{M}_\odot$ and $\sim 0.53~\rm{M}_\odot$, respectively, using a mass of $\sim 1.83$--$2.14~\rm{M}_\odot$ for T Tau N \citep{loinard2007}. Recent near-infrared polarimetry suggests an Sa-Sb circumbinary disk (CBD) with a size of $\sim 150~\text{mas}$ and a position angle of $\sim 30~\text{deg}$ \citep{yang2018}. This would imply that T Tau Sb is currently moving along its orbit around Sa through the Sa-Sb CBD plane and should therefore suffer noticeable extinction for the first time since its discovery. Indeed, T Tau Sb has recently been dimming in the near-infrared \citep{schaefer2020}, and \citet{koehler2020} proposed that this dimming is caused by Sb passing through the Sa-Sb CBD plane, based on the J--K-band photometry of the system.

In this paper, we present new mid-infrared ($\sim 10~\text{\textmu m}$) photometry of all three components of the T Tau system obtained with the VLT Imager and Spectrometer for mid-IR-New Earths in the Alpha Centauri Region instrument (VISIR-NEAR) using kernel phase interferometry. Our observations were conducted in December 2019 when the southern binary was separated by only $\sim 64~\text{mas}$, which corresponds to $\sim 0.2~\lambda/D$ at the longest observed wavelengths. Our data allow us to study the brightness of the individual components on and off the $9.7~\text{\textmu m}$ silicate feature as well as their variability. The silicate feature is interesting since it contains information about the nature of the circumstellar dust and about the orientation and geometry of the CSDs in this ever-fascinating system.

\section{Observations}
\label{sec:observations}

We observed T~Tau with the VISIR-NEAR instrument of the European Southern Observatory's (ESO) VLT at the summit of Cerro Paranal in Chile. VISIR-NEAR is mounted on the Cassegrain focus of Unit Telescope 4 (UT4) and consists of the VISIR instrument \citep{lagage2004} plus the NEAR upgrade \citep{kaeufl2018}, which enables mid-infrared imaging with extreme adaptive optics. Our observations are part of the VISIR-NEAR science demonstration program 60.A-9107(E), PI M.~Kasper, and are summarized in Table~\ref{tab:obslog}.

We acquired data for both T~Tau and a PSF reference (HD~27639) in pupil-tracking mode, at wavelengths of $8.98~\text{\textmu m}$ (filter name ARIII), $9.81~\text{\textmu m}$ (SIV\_1), $10.71~\text{\textmu m}$ (SIV\_2), and $12.80~\text{\textmu m}$ (NEII). The sky offset between T Tau and the PSF reference is $\sim 1.3~\text{deg,}$ and the difference in average airmass is $\sim 0.04$. Atmospheric turbulence was removed almost entirely by the adaptive optics facility \citep[AOF,][]{stroebele2006} of UT4, leaving quasi-static aberrations (such as chromatic aberrations of the filters) as the dominant systematic error. These errors can be identified and calibrated by our kernel phase analysis.

\begin{table*}
\caption{Observing log for program 60.A-9107(E). All data were taken on December 16, 2019. Reported are the number of exposures per chopping position (NDIT), the exposure time (DIT), the number of chopping cycles ($N_\text{chop}$), the chopping frequency ($\nu_\text{chop}$), the number of frames after averaging ($N_\text{avg}$), and the number of good frames ($N_\text{good}$).}
\label{tab:obslog}
\centering
\begin{tabular}{c c c r c c c c c c}
\hline\hline
Start time (UT) & Object & Filter & $\lambda$ [$\text{\textmu m}$] & NDIT & DIT [ms] & $N_\text{chop}$ & $\nu_\text{chop}$ [Hz] & $N_\text{avg}$ & $N_\text{good}$ \\
\hline
02:01:24 & T Tau & NEII & 12.80 & 8 & 25 & 66 & 2.222 & 10 & 10 \\
02:08:18 & T Tau & SIV\_1 & 9.81 & 9 & 22 & 75 & 2.273 & 12 & 6 \\
02:16:51 & T Tau & SIV\_2 & 10.71 & 9 & 22 & 75 & 2.273 & 12 & 12 \\
02:25:28 & T Tau & ARIII & 8.98 & 6 & 80 & 40 & 0.893 & 15 & 7 \\
03:05:48 & HD27639 & NEII & 12.80 & 8 & 25 & 66 & 2.222 & 4 & 3 \\
03:08:58 & HD27639 & SIV\_1 & 9.81 & 9 & 22 & 75 & 2.273 & 4 & 2 \\
03:12:26 & HD27639 & SIV\_2 & 10.71 & 9 & 22 & 75 & 2.273 & 8 & 6 \\
03:28:05 & HD27639 & ARIII & 8.98 & 6 & 80 & 40 & 0.893 & 4 & 2 \\
\hline
\end{tabular}
\end{table*}

\section{Data reduction}
\label{sec:data_reduction}

At the time of our observations, the southern binary of the T Tau triple system was separated by $\sim 64~\text{mas,}$ which is equivalent to $0.28$, $0.26$, $0.24$, and $0.20~\lambda/D$ in the ARIII, SIV\_1, SIV\_2, and NEII filters, respectively. Such small separations are very challenging for high-resolution single-dish imaging. Since VISIR-NEAR is mounted at the UT4, which also hosts the AOF, the instrument achieves extremely high Strehl ratios in the mid-infrared. This makes our data set ideal for an analysis with the kernel phase technique, which models a single-dish telescope as an interferometer during the post-processing of the images. The kernel phase technique relies on a linear relationship between the pupil plane phase and the Fourier phase of the images, possible only in the high-Strehl regime. With the known astrometry of the T Tau triple system from observations at shorter wavelengths, we can use the kernel phase technique to obtain its photometry in the mid-infrared.

\subsection{Kernel phase technique}
\label{sec:kernel_phase_technique}

The kernel phase technique was developed by \citet{martinache2010}, who was able to achieve an angular resolution of $\sim 0.6~\lambda/D$ on the known low-contrast binary GJ~164 with the Near-Infrared Camera and Multi-Object Spectrometer (NICMOS) on the Hubble Space Telescope (HST). Later, \citet{pope2013} used the technique to detect brown dwarf companions, also with HST/NICMOS, and \citet{pope2016} and \citet{sallum2019} showed that kernel phase should outperform sparse aperture masking under appropriate seeing conditions longward of $\sim 3~\text{\textmu m}$ from the ground given the reduced sensitivity to sky background noise. Recently, \citet{kammerer2019} used the kernel phase technique to detect eight stellar companions in an archival VLT/Nasmyth Adaptive Optics System (NAOS) near-infrared imager and spectrograph (NACO) high-contrast imaging survey, two of which are below the classical diffraction limit at angular separations of $\sim 0.8$ and $\sim 1.2~\lambda/D$.

The kernel phase technique achieves its high resolution by making use of the superior calibration capabilities of the Fourier plane phase (i.e., the phase of the Fourier transform of the image) in the high-Strehl regime. Here, one can linearize the expression of the Fourier plane phase $\phi$ as a function of the pupil plane phase $\varphi$, that is,
\begin{equation}
    \label{eqn:fourier_phase}
    \phi(\varphi) = \bm{R}^{-1} \cdot \bm{A} \cdot \varphi+\phi_\text{obj}+\mathcal{O}(\varphi^3),
\end{equation}
where $\bm{A}$ is the matrix of a linear map between a pair of pupil plane positions and their corresponding spatial frequency, $\bm{R}$ is a diagonal matrix encoding the redundancy of the spatial frequencies, and $\phi_\text{obj}$ is the Fourier plane phase of the object intensity distribution (i.e., of the astronomical source). The first step was to model the single-dish telescope as an interferometer by discretizing the telescope pupil onto a grid of so-called subapertures in order to find a simple representation of matrix $\bm{A}$, and our VISIR-NEAR pupil model is shown in Figure~\ref{fig:pupil_model}. In the high-Strehl regime, the higher order pupil plane phase noise $\mathcal{O}(\varphi^3)$ is negligible and one can multiply Equation~\ref{eqn:fourier_phase} with the kernel $\bm{K}$ of $\bm{R}^{-1} \cdot \bm{A}$ to obtain the kernel phase $\theta$, namely,\begin{equation}
    \theta = \bm{K} \cdot \phi = \underbrace{\bm{K} \cdot \bm{R}^{-1} \cdot \bm{A} \cdot \varphi}_{=0}+\bm{K} \cdot \phi_\text{obj}+\mathcal{O}(\varphi^3) \approx \theta_\text{obj}.
\end{equation}
This is a very powerful finding as it means that, to the second order, the kernel phase measured in the image is directly equivalent to the kernel phase of the astronomical source.

\subsection{Basic cleaning}
\label{sec:basic_cleaning}

Table~\ref{tab:obslog} shows an observing log for program 60.A-9107(E). Each chopping cycle consists of two chopping positions, at each of which NDIT exposures are saved and averaged into a single frame by the instrument control software. Then, for each of the $N_\text{chop}$ chopping cycles, we subtracted the two frames from the different chopping positions from each other in order to remove the mid-infrared background and averaged that over the $N_\text{chop}$ background-subtracted frames; doing so, we obtained one cleaned frame for each of the $N_\text{avg}$ observations. We note that we also nodded along the chopping axis during the observations but performed no nodding subtraction here for the sake of simplicity.

Next, we found the brightest PSF in the field-of-view (FOV) by looking for the brightest pixel in the median filtered (3 pixel) frame and cropped the frames to a size of 128 $\times$ 128 pixels around the brightest PSF, which corresponds to an FOV of 5.792 arcsec${}^2$. A more sophisticated sub-pixel re-centering is performed at a later stage (see Section~\ref{sec:kernel_phase_extraction_and_calibration}).

The detector position angle of each frame was computed as the arithmetic mean of the ``POSANG'' and ``POSANG END'' fits header keywords. This approximation is feasible since the observing times are short and the field rotation per fits file is only on the order of $0.25~\text{deg}$. Unfortunately, the detector position angle computed from the fits header keywords does not represent the true position angle of the detector. There is an additional rotation offset $\vartheta$ that we determined from VISIR-NEAR data of the nearby and well-known binary system $\alpha$ Centauri. With the known astrometry of this binary system, we obtain a rotation offset of $\vartheta = 94.02 \pm 0.10~\text{deg}$ and a pixel scale of $45.25 \pm 0.10~\text{mas}$. The detector position angle and the rotation offset were both measured clockwise (i.e., west of north), while the position angle of the companions was measured counterclockwise (i.e., east of north), in accordance with the standard convention.

\subsection{Kernel phase extraction and calibration}
\label{sec:kernel_phase_extraction_and_calibration}

\begin{figure*}
\centering
\includegraphics[width=0.34\textwidth]{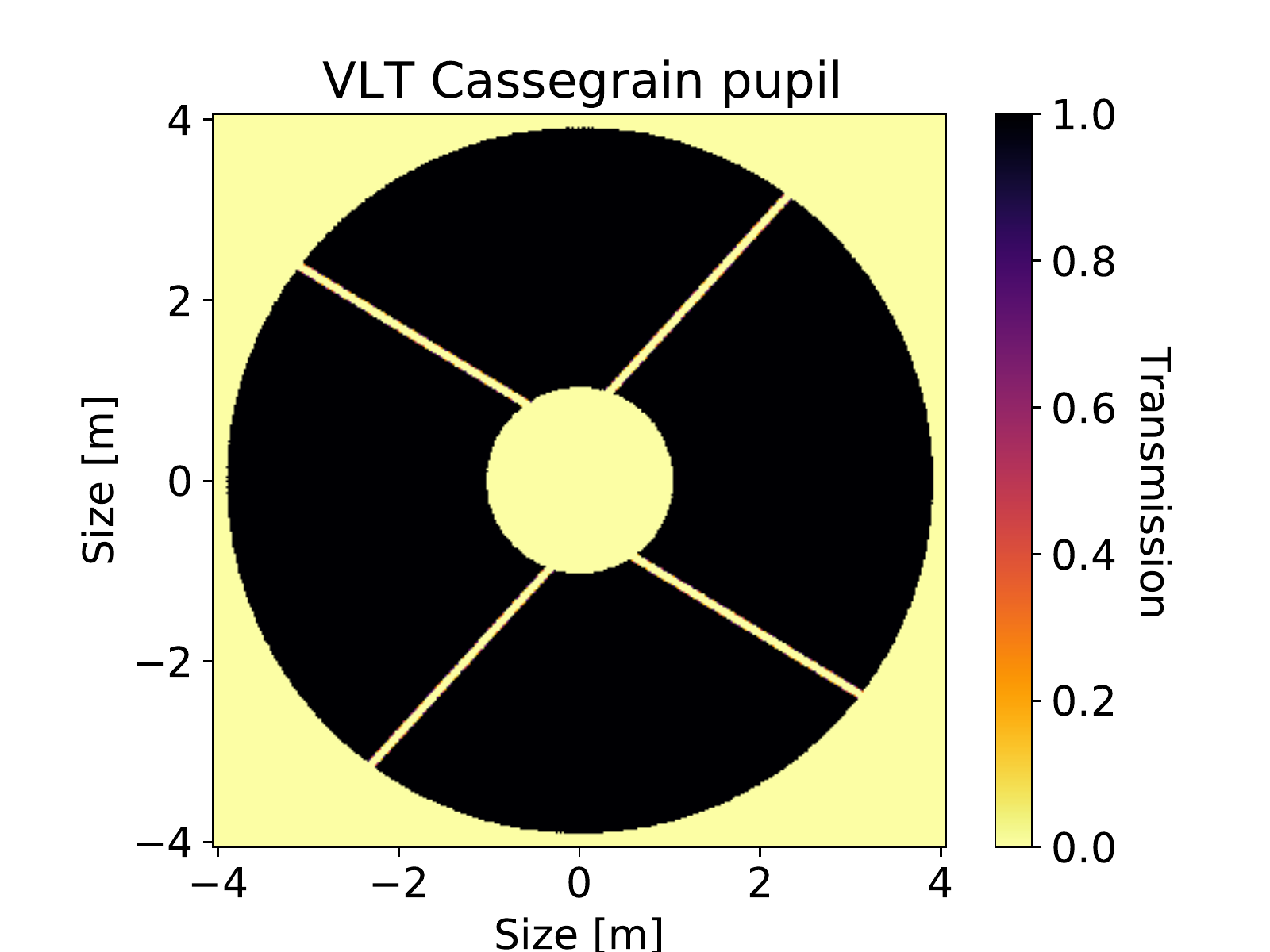}
\includegraphics[width=0.64\textwidth]{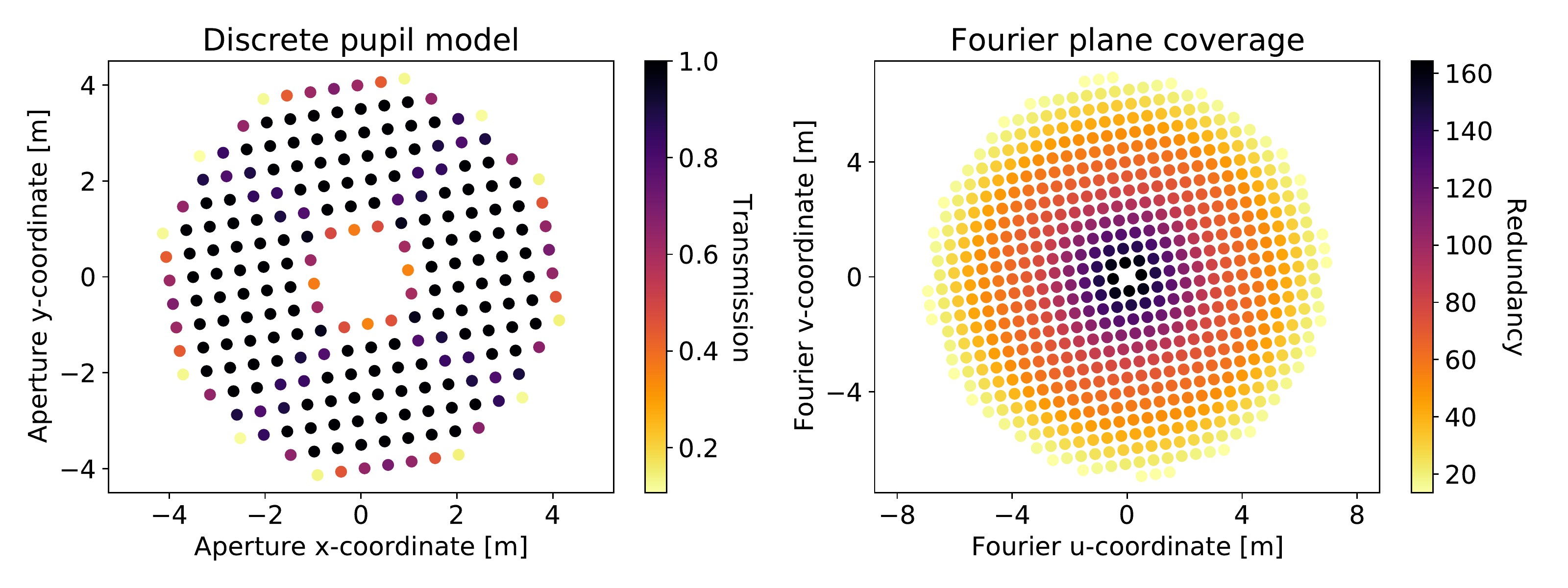}
\caption{VISIR-NEAR pupil model discretized onto a grid of subapertures for the kernel phase analysis. Left: VLT Cassegrain pupil with an extended central obscuration due to the M3 baffle and four thin M2 support struts (``spiders''), rotated by $81.8~\text{deg}$ clockwise. Middle: Our discrete pupil model consisting of 224 subapertures, spanning 310 distinct baselines with 88 kernel phases. Right: Fourier plane coverage of our discrete pupil model.} 
\label{fig:pupil_model}
\end{figure*}

To extract the kernel phase from the cleaned frames, we used the \texttt{XARA}\footnote{\url{https://github.com/fmartinache/xara/}} package \citep{martinache2010,martinache2013}. \texttt{XARA} requires a discrete pupil model of the VISIR-NEAR instrument in order to compute the kernel matrix $\bm{K}$. Figure~\ref{fig:pupil_model} shows the VLT Cassegrain pupil (left-hand panel), our discrete pupil model consisting of 224 subapertures placed on a grid with a spacing of $0.5~\text{m}$ (middle panel), and the Fourier plane coverage of our discrete pupil model (right-hand panel). The use of subapertures with a continuous transmission between $0$ and $1$ (referred to as gray apertures) helps to significantly minimize residual calibration errors  \citep{martinache2020}. We note that the pupil is rotated by $81.8~\text{deg}$ clockwise (west of north, see Figure~\ref{fig:pupil_model}) with respect to the detector and we aligned the grid of subapertures with the pupil in order to obtain a symmetric pupil model and reduce calibration errors \citep[see][]{martinache2020}. We determined this rotation from VISIR-NEAR data of $\alpha$ Centauri, which feature bright diffraction artifacts from the secondary mirror (M2) support spiders.

\texttt{XARA} first masks the cleaned frames with an exponential windowing function $m$ (sometimes called super-Gaussian),
\begin{equation}
    m(x, y) = \exp{\left(-\left(\frac{\sqrt{(x-x_0)^2+(y-y_0)^2}}{r}\right)^4\right)}
,\end{equation}
with a radius $r$ of 50 pixels ($\sim 2.263~\text{arcsec}$) in order to minimize edge effects during the Fourier transform, where $x$ and $y$ are the x- and y-coordinates of the frame and $x_0 = 64$ and $y_0 = 64$ are the coordinates of the center of the frame. This is equivalent to a Gaussian smoothing in the Fourier plane, and a value of 50 pixels is large enough to not interfere with the spatial frequencies measured by the VLT. It then extracts the Fourier plane phase $\phi$ from the cleaned (dither-subtracted) frames using a discrete Fourier transform at the uv-positions corresponding to the subapertures of our pupil model. Finally, it performs sub-pixel re-centering in the complex visibility space by multiplying the Fourier phase $\phi$ with a wedge function and computes the kernel phase $\theta = \bm{K} \cdot \phi$. This wedge function is determined by computing the photocenter of the brightest PSF on an iteratively shrinking window.

Furthermore, we estimated the kernel phase covariance $\bm{\Sigma}_\theta$ based on the pixel-to-pixel background variance in the cleaned frames. Therefore, we applied the basis transform 
\begin{equation}
    \bm{\Sigma}_\theta = \bm{B} \cdot \bm{\Sigma}_d \cdot \bm{B}^T,
\end{equation}
where
\begin{equation}
    \bm{B} = \bm{K} \cdot \frac{\mathrm{Im}(\bm{F})}{|\bm{F} \cdot \bm{d} \cdot g \cdot \bm{m}|}
\end{equation}
and $\bm{\Sigma}_d$ is the variance frame \citep[see][]{kammerer2019}. Here, we obtain the variance frame directly from the frame itself, namely,
\begin{equation}
    \Sigma_d = \mathrm{var}(\bm{d} \cdot g) \cdot \bm{m}^2,
\end{equation}
where $\bm{d}$ is the frame itself and $g$ is the gain in photoelectrons per detector count. We note that we masked out the center of the frame with a circular aperture of 80 pixels ($3.620~\text{arcsec}$) in diameter in order to avoid confusion from the PSF of T Tau.

\begin{figure*}
\centering
\includegraphics[width=0.49\textwidth]{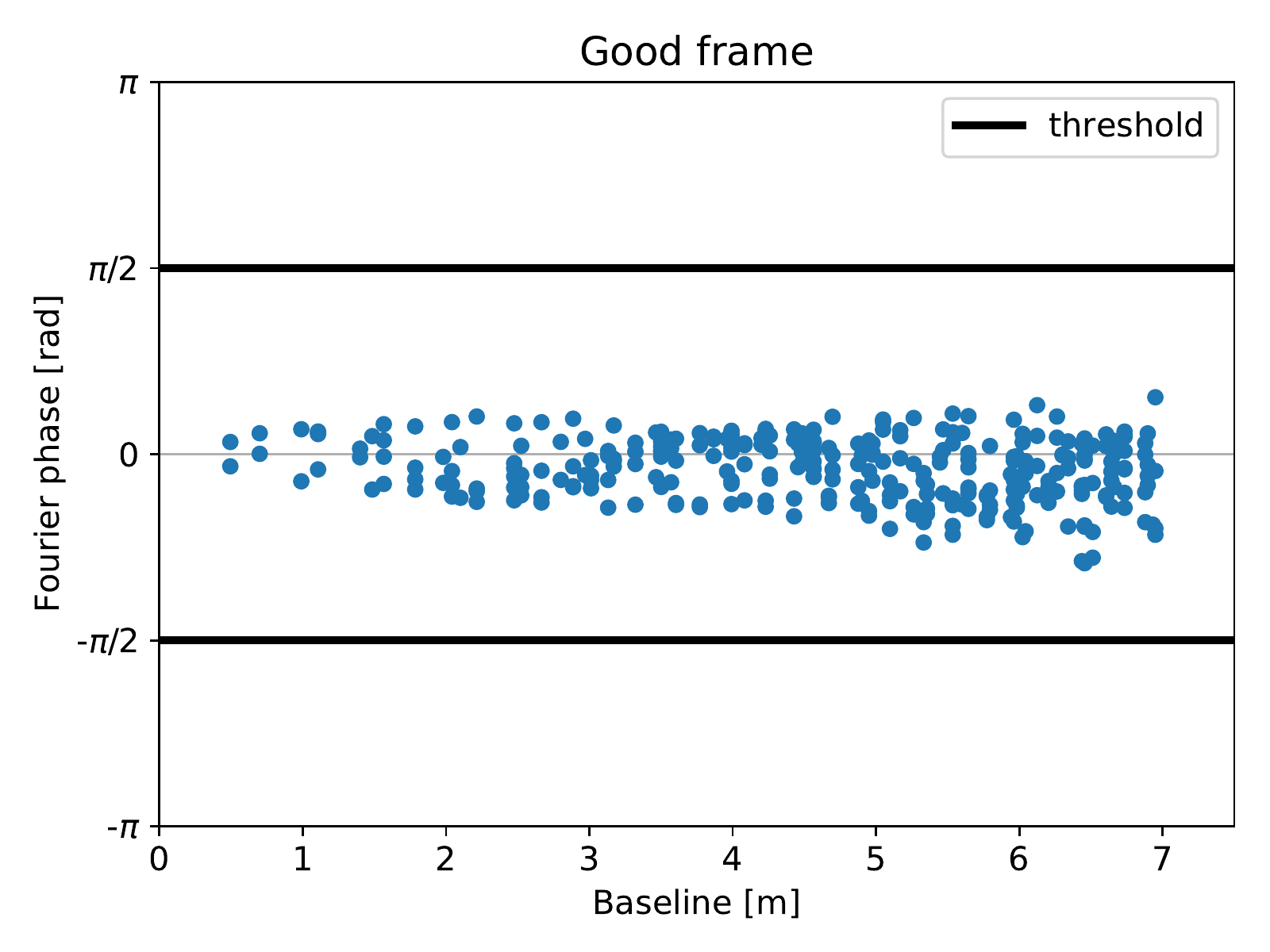}
\includegraphics[width=0.49\textwidth]{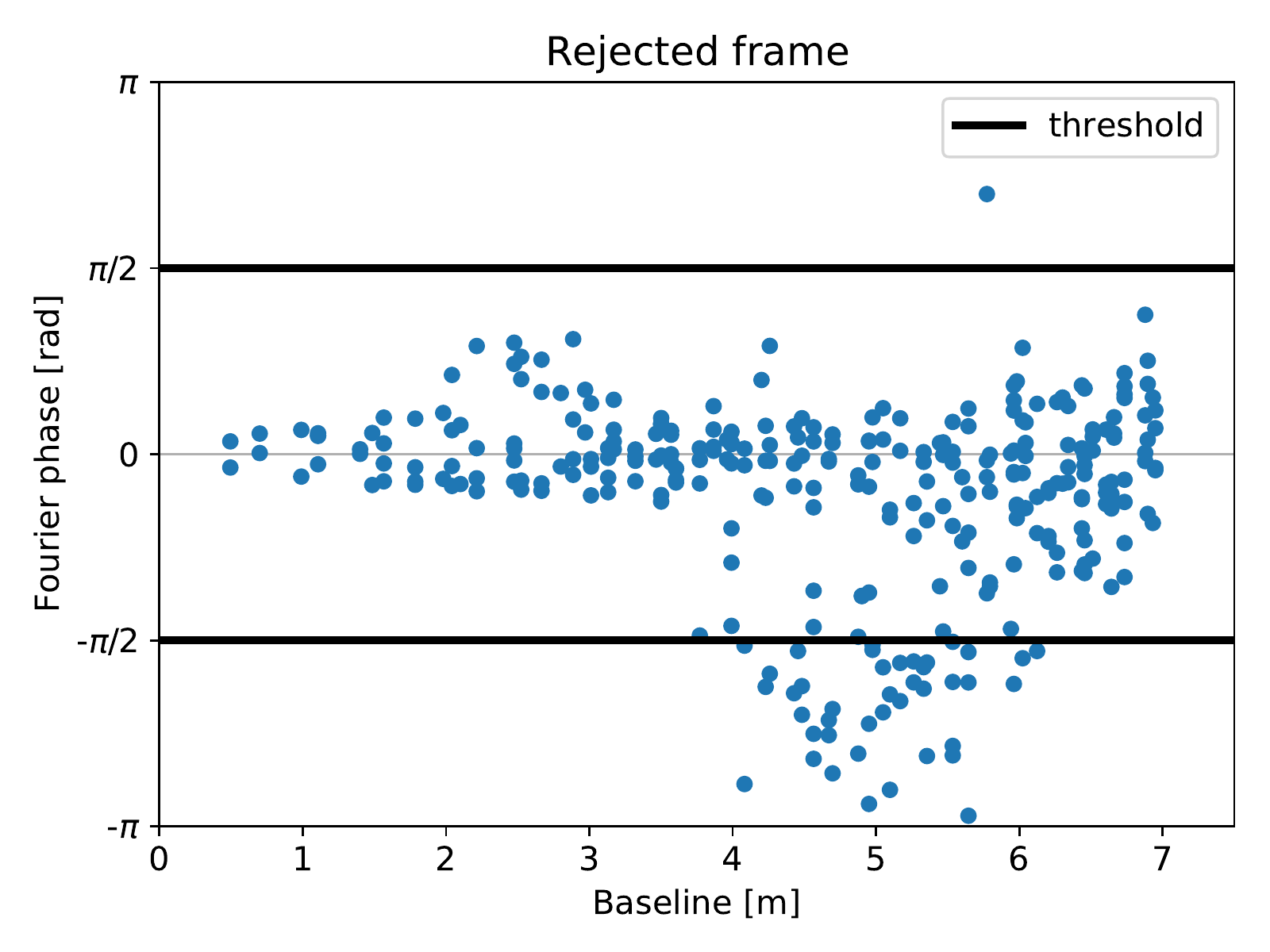}
\caption{Frame selection criterion used to reject bad frames that are not suitable for a kernel phase analysis. Left: Fourier phase $\phi$ as a function of baseline for an example frame taken with the SIV\_1 filter that passes our selection criterion ($|\phi| < 90~\text{deg}$). Right: Same, but for an example frame that does not pass our selection criterion due to seeing-driven aberrations.}
\label{fig:frame_selection}
\end{figure*}

Before continuing with the calibration and the model fitting, we made a frame selection based on the absolute Fourier phase $|\phi|$. We rejected any frame whose absolute Fourier phase $|\phi|$ exceeds 90 deg on at least one of the 310 baselines, that is to say, vectors connecting two subapertures (see Figure~\ref{fig:frame_selection}). This threshold is chosen empirically and to be sufficiently high so that frames with a high-Strehl PSF of T Tau are kept (we note that T Tau is a low-contrast triple system and therefore has a nonzero Fourier phase) but that frames suffering seeing-driven aberrations are rejected. For frames whose absolute Fourier phase $|\phi|$ exceeds 90 deg, higher order pupil plane phase errors are large and would add significant systematic errors. The number of good frames $N_\text{good}$ obtained for T Tau and the PSF reference for each filter can be found in Table~\ref{tab:obslog}. The number of rejected frames can be obtained as the difference between $N_\text{avg}$ and $N_\text{good}$.

In order to calibrate the kernel phase, we subtracted the average of the kernel phase measured on the calibrator from the kernel phase measured on the science target \citep[see e.g.,][]{martinache2010}. The corresponding kernel phase covariance then writes as\begin{equation}
    \bm{\Sigma}_\theta = \bm{\Sigma}_{\theta,\text{sci}}+\frac{1}{N_\text{good}^2}\sum_{n=1}^{N_\text{good}}\bm{\Sigma}_{\theta,\text{cal},n}.
\end{equation}

\subsection{Model fitting}
\label{sec:model_fitting}

With the kernel phase $\theta$ and its covariance $\bm{\Sigma}_\theta$, we can perform model fitting based on likelihood maximization or $\chi^2$ minimization. For this, we used a model of $N$ point sources,
\begin{equation}
    \label{eqn:model}
    \theta_\text{mod} = \bm{K} \cdot \mathrm{arg}\left(\sum_{n=1}^N\xi_nc_n\exp\left(-2\pi i\left(\frac{\Delta_{\text{RA},n}u}{\lambda}+\frac{\Delta_{\text{DEC},n}v}{\lambda}\right)\right)\right),
\end{equation}
where $\xi_n$ is the relative flux of the n-th component (normalized to the total flux), $c_n$ is the complex visibility of the n-th component, $\Delta_{\text{RA},n}$ and $\Delta_{\text{DEC},n}$ are the sky offset of the n-th component from the center of the frame, $u$ and $v$ are the Fourier u- and v-coordinates of our pupil model, and $\lambda$ is the observing wavelength. In case of unresolved point sources (such as is the case for our observations), the complex visibility $c_n = 1$.

The N-band flux of both T Tau Sa and Sb is entirely dominated by emission from their CSDs. Both CSDs are tidally truncated by the binary's orbit, which has a semi-major axis of $\sim 12.5~\text{au}$ and an eccentricity of $\sim 0.55$ \citep{schaefer2020}. For such orbital parameters, \citet{artymowicz1994} predict a ratio of the disk truncation radius to the semimajor axes that would be on the order of $\sim 0.25$ ($= 3.1~\text{au}$) for T Tau Sa and $\sim 0.15$ ($= 1.9~\text{au}$) for Sb. This corresponds to angular sizes of 21~mas and 13~mas, respectively, for the CSDs of T Tau Sa and Sb and is significantly smaller than the diffraction limit at our shortest observing wavelength of $\sim 230~\text{mas}$. In addition, the CSDs get warmer and brighter toward the inner disk rim, with radii of a few tenths of an au ($\lessapprox 3~\text{mas}$). Therefore, we can safely assume that T Tau Sa and Sb are spatially unresolved in our data.

In the following, we used a more intuitive representation $f_n$ of the relative flux of the individual components by normalizing it to the flux of the brightest component so that $f_1 = 1$ and $f_{n>1} \leq 1$. The $\xi_n$ are related to the $f_n$ according to
\begin{equation}
    \xi_n = \frac{f_n}{\sum_{n=1}^Nf_n}.
\end{equation}

If we are in the high-contrast regime ($\xi_{n>1} \ll 0.5$) and only fit for two components, the model can be simplified by linearization \citep{kammerer2019} and by putting the brighter component into the center of the frame ($\Delta_{\text{RA},1} = \Delta_{\text{DEC},1} = 0$). Then, we can solve analytically for the relative companion flux,
\begin{equation}
    \xi_2 = \frac{\theta_\text{ref}^T \cdot \bm{\Sigma}_\theta^{-1} \cdot \theta}{\theta_\text{ref}^T \cdot \bm{\Sigma}_\theta^{-1} \cdot \theta_\text{ref}}
,\end{equation}
on an RA-DEC grid, where $\theta_\text{ref}$ is the binary model (Equation~\ref{eqn:model} for $N = 2$) evaluated on an RA-DEC grid and for a small relative reference flux $\xi_{2,\text{ref}} \ll 0.5$ \citep[see][]{kammerer2019}. We note that with the T Tau system, we are not in the high-contrast regime where this simplification holds. Nevertheless, such a grid search allows us to find a good first guess for the relative companion flux and its position, which we can then use as an initial position for a least-squares or a Markov chain Monte Carlo (MCMC) algorithm. Such algorithms directly minimize the $\chi^2$ of the model, that is,\begin{equation}
    \chi^2 = R^T \cdot \bm{\Sigma}_\theta^{-1} \cdot R,
\end{equation}
where $R = \theta-\theta_\text{mod}$ is the residual kernel phase signal between the data and the model, so that no linearization is necessary.

\section{Results and analysis}
\label{sec:results_and_analysis}

Our goal is to determine the flux of T Tau Sa and Sb relative to T Tau N in all four filters. Therefore, we first performed a binary model fit in order to find the combined flux of T Tau S (i.e., T Tau Sa and Sb) relative to T Tau N. Then, we performed a triple model fit using half of the combined relative flux of T Tau S as initial values for the fluxes of T Tau Sa and Sb relative to T Tau N (making the simple assumption that T Tau Sa and Sb are equally bright; the least-squares or MCMC algorithm will then find the true flux ratio). We note that for the NEII filter we proceeded slightly differently since T Tau S, and not T Tau N, is the brightest component (see Section~\ref{sec:neii_data}). Finally, we created mock data with a range of different fluxes for T Tau Sa and Sb relative to T Tau N and repeated the triple model fits in order to demonstrate that our kernel phase model fitting technique correctly reproduces the mock data.

\subsection{Binary model fits }
\label{sec:binary_model_fits}

\begin{figure*}
\centering
\includegraphics[width=0.8\textwidth]{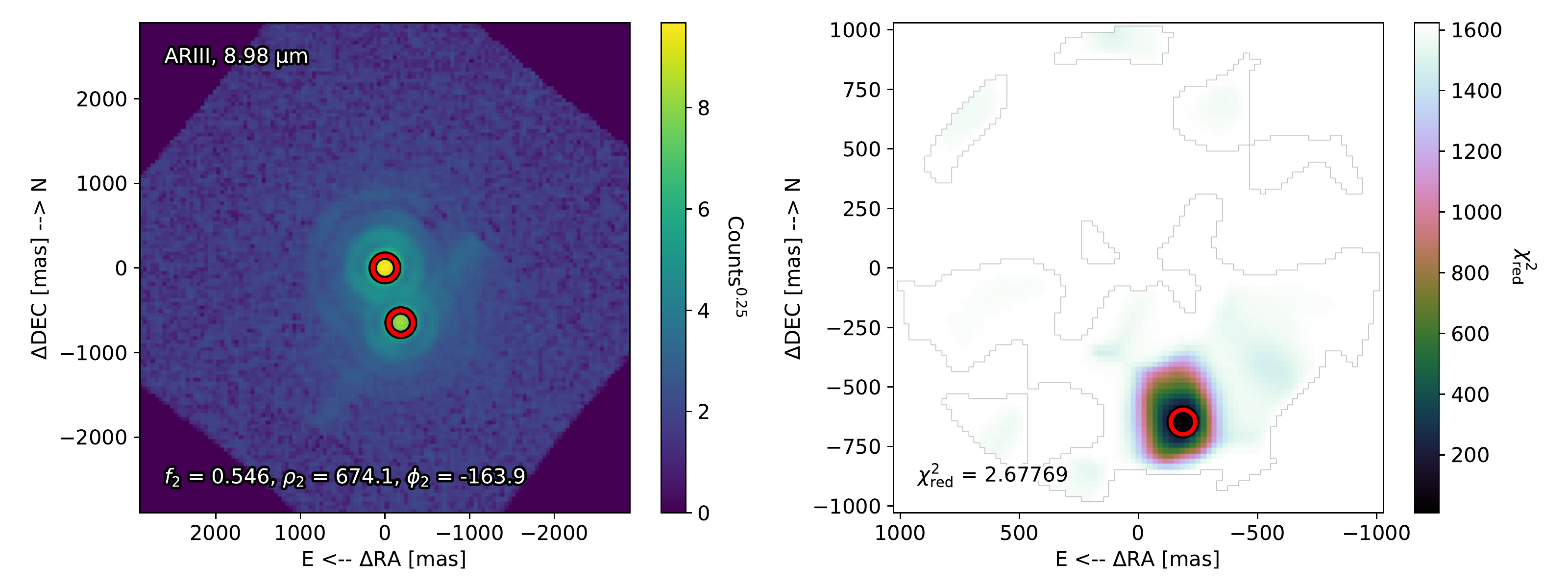}
\includegraphics[width=0.8\textwidth]{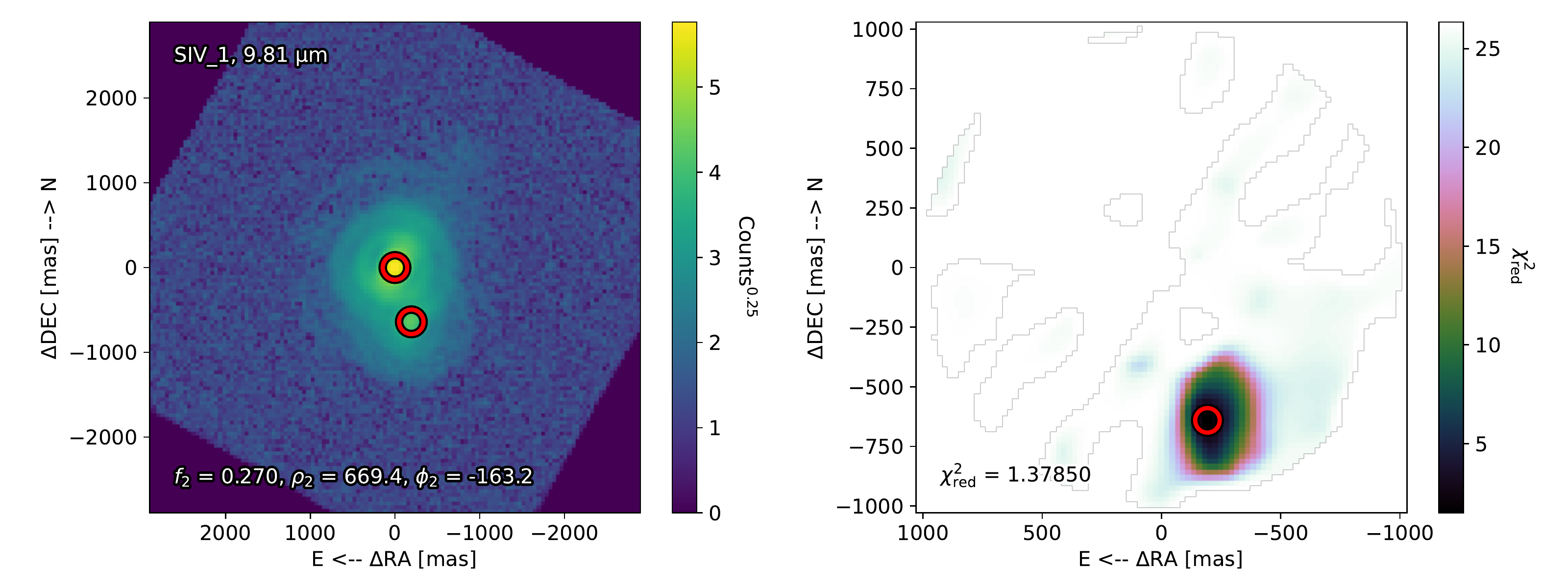}
\includegraphics[width=0.8\textwidth]{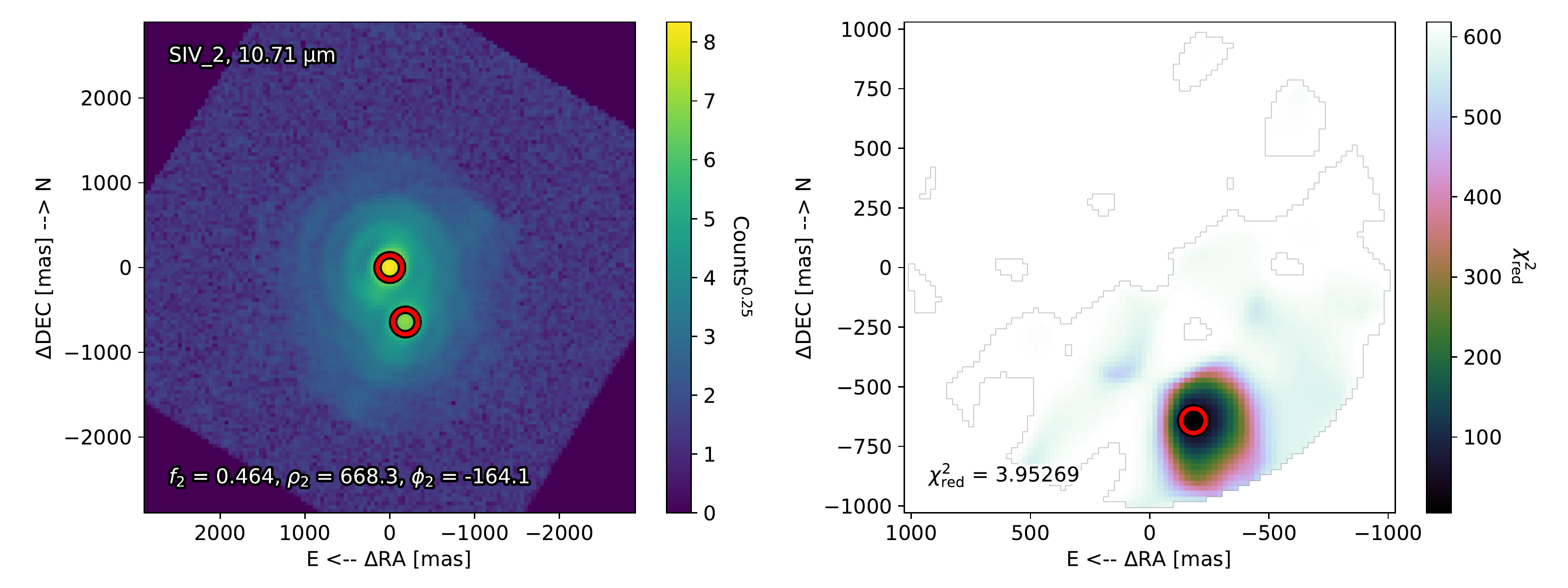}
\includegraphics[width=0.8\textwidth]{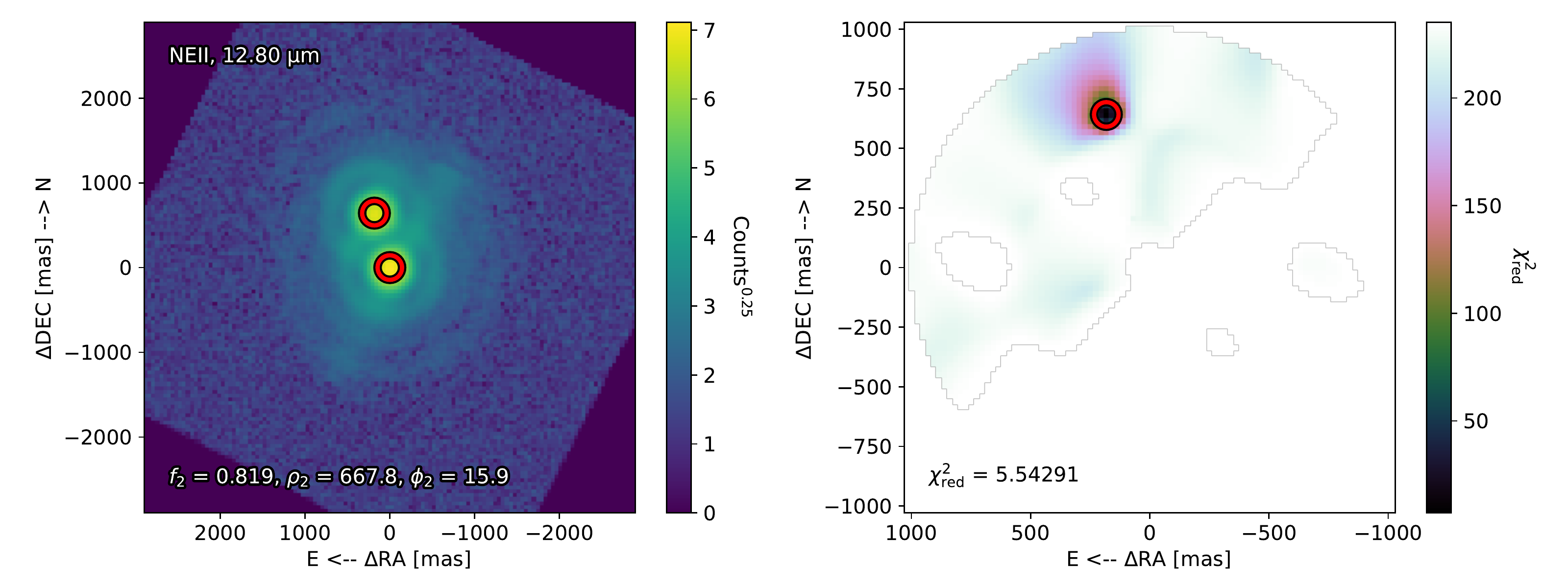}
\caption{De-rotated and co-added VISIR-NEAR images of the T Tau triple system and the two-component fit (left-hand column, red circles) and the corresponding maps of the reduced $\chi^2$ (right-hand column) in the ARIII, SIV\_1, SIV\_2, and NEII filters (from top to bottom). In the right-hand panels, the red circle highlights the best fit position for the companion (here, the companion is assumed to be the combined PSF of T Tau Sa and Sb). The relative flux $f_2$, the angular separation $\rho_2$ [mas], and the position angle $\phi_2$ [deg] of the best fit are printed in the left-hand panels, and its reduced $\chi^2$ is printed in the right-hand panels. We note that the NEII data are centered on T Tau S, which is the brighter component at $12.80~\text{\textmu m}$.}
\label{fig:binary_model_fits}
\end{figure*}

Figure~\ref{fig:binary_model_fits} shows the results of our binary model fits to the T Tau system, with both position and flux left as free parameters, yielding a combined flux of T Tau S relative to T Tau N of $0.546 \pm 0.015$ (ARIII), $0.270 \pm 0.064$ (SIV\_1), $0.464 \pm 0.051$ (SIV\_2), and $1.217 \pm 0.047$ (NEII). The position uncertainties are on the order of $\sim 10~\text{mas}$, constrained by the astrometric precision of the kernel phase technique; the best fit position varies slightly from filter to filter, also due to the changing flux ratio between T Tau Sa and Sb, which shifts the photocenter of the combined southern component. We note that the NEII data are centered on T Tau S, which is the brighter component at $12.80~\text{\textmu m}$. This already reveals some differences compared to the values reported by \citet{skemer2008}, who found the combined southern component to be brighter than the northern component at $8.7~\text{\textmu m}$ ($\sim$ ARIII). At $10.55~\text{\textmu m}$ ($\sim$ SIV\_2), our result of a flux ratio of $0.464 \pm 0.051$ between the southern and the northern component is more similar to the value of $\sim 0.57$ reported by \citet{skemer2008}, and we find the combined southern component to only be brighter than the northern component at the longest wavelength ($12.80~\text{\textmu m}$).

Since T Tau N is fainter than T Tau S at $12.80~\text{\textmu m}$ (NEII), the kernel phase analysis is slightly more complicated. Therefore, we only report on the analysis of the ARIII, SIV\_1, and SIV\_2 data in the following and treat the NEII data separately in Section~\ref{sec:neii_data}. Furthermore, the reduced $\chi^2$ of our binary model fits is larger than one, meaning that we are underestimating the errors. This is in agreement with previous studies that have shown that the kernel phase technique is dominated by systematic errors, usually yielding a reduced $\chi^2$ between $\sim 1$--$10$ \citep{martinache2020,laugier2020,kammerer2019,ireland2013}.

\subsection{Triple model fits}
\label{sec:triple_model_fits}

For the triple model fits, we used the known offsets of T Tau Sa and Sb relative to T Tau N, which are (-648.1~mas N, -188.5~mas E) for T Tau Sa and (-595.6 mas N, -151.4~mas E) for T Tau Sb. We obtained these values from extrapolating the orbits of the T Tau system from \citet{koehler2016}, which were obtained from VLT/SPHERE astrometry, for the date of our observations. Their accuracy is $\sim 1~\text{mas}$.

\begin{figure*}
\centering
\includegraphics[width=0.5\textwidth]{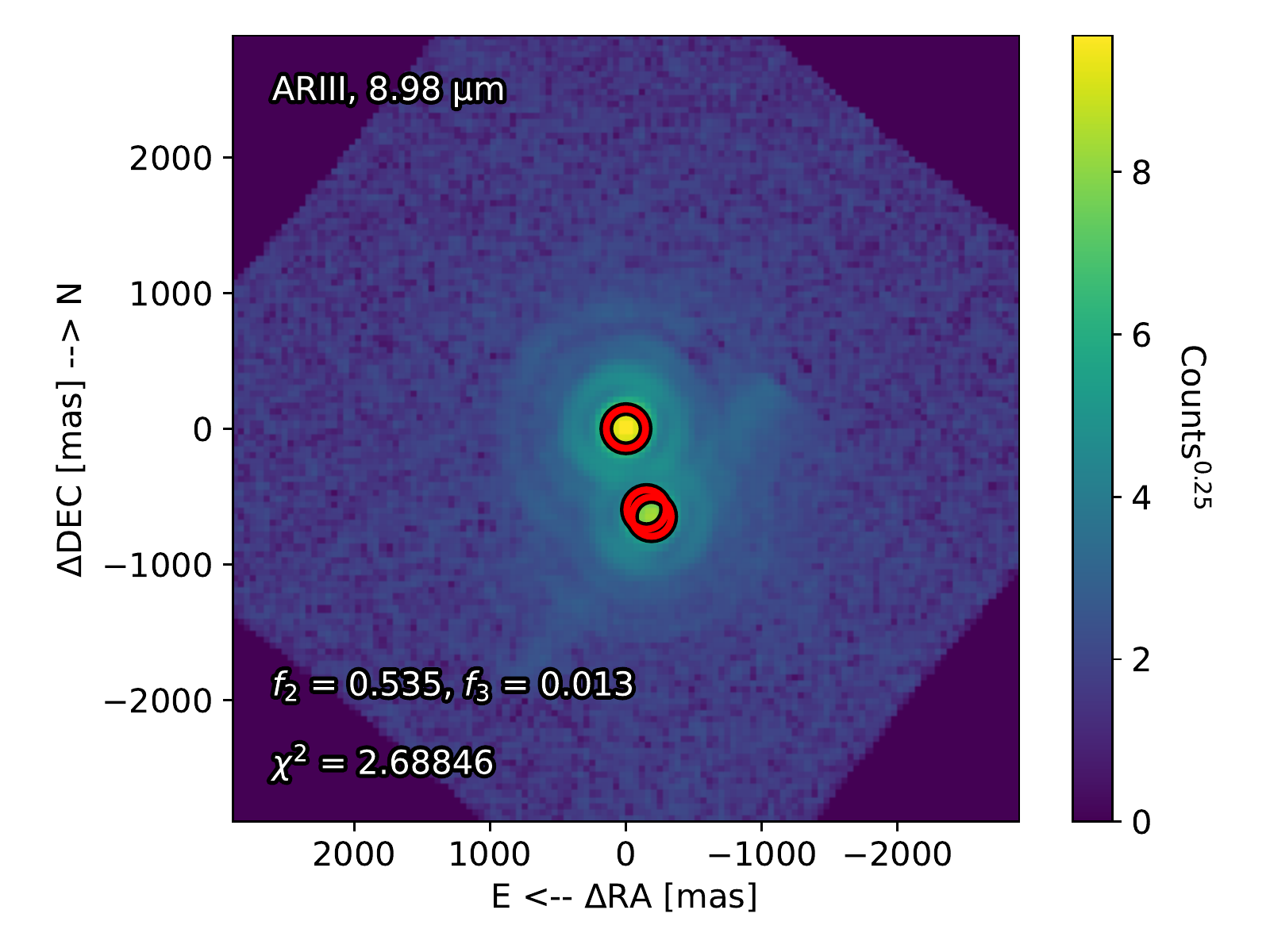}
\includegraphics[width=0.4\textwidth]{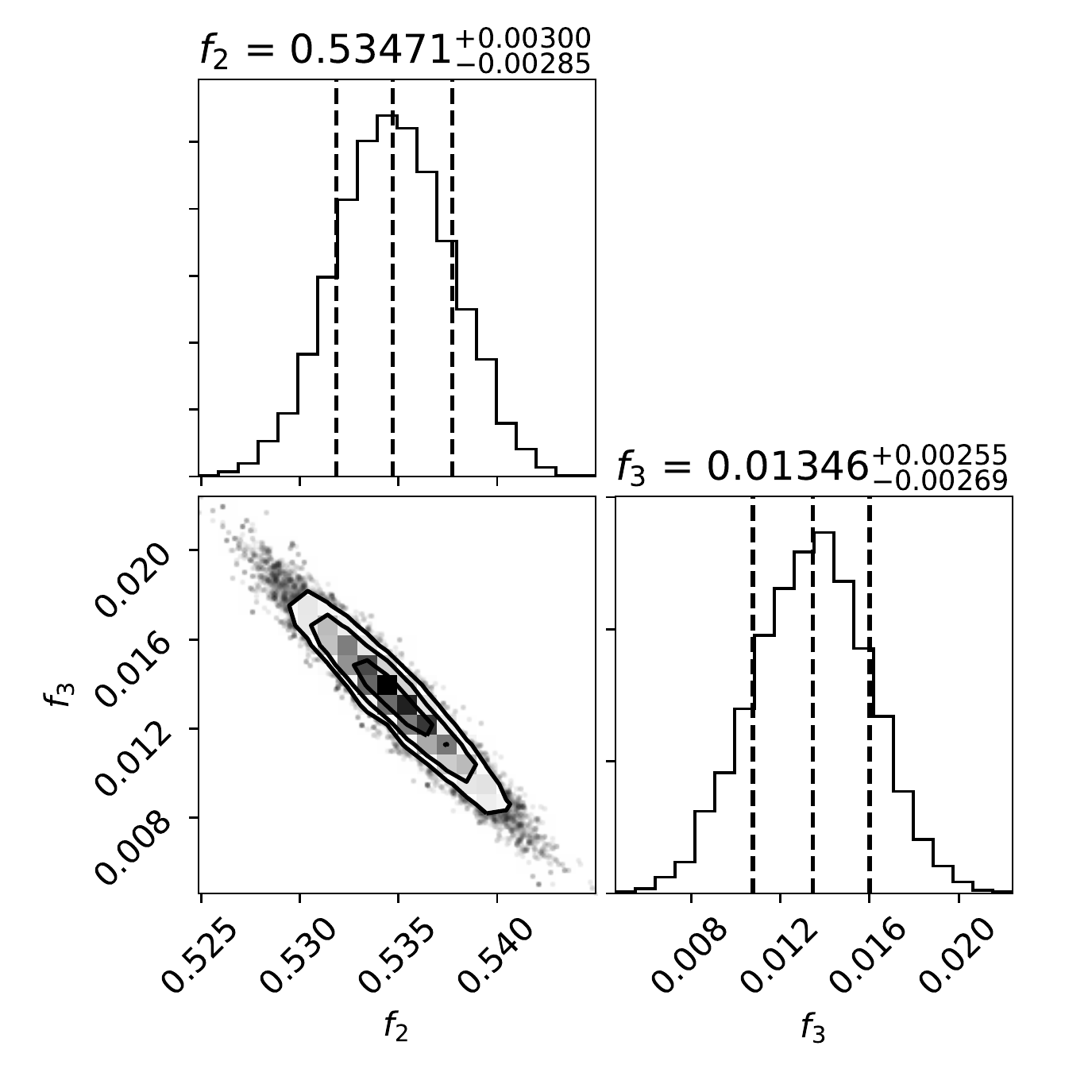}
\includegraphics[width=0.5\textwidth]{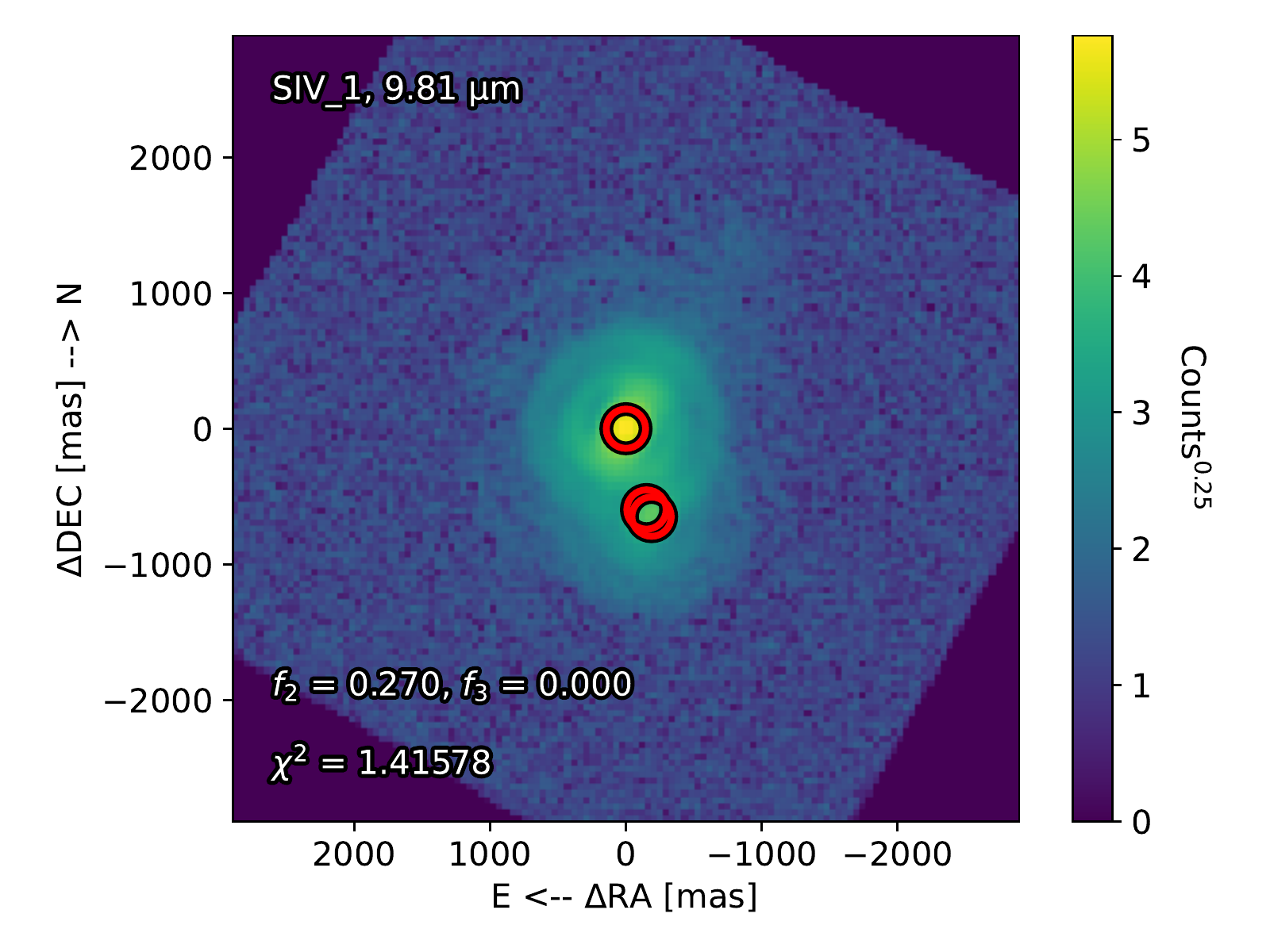}
\includegraphics[width=0.4\textwidth]{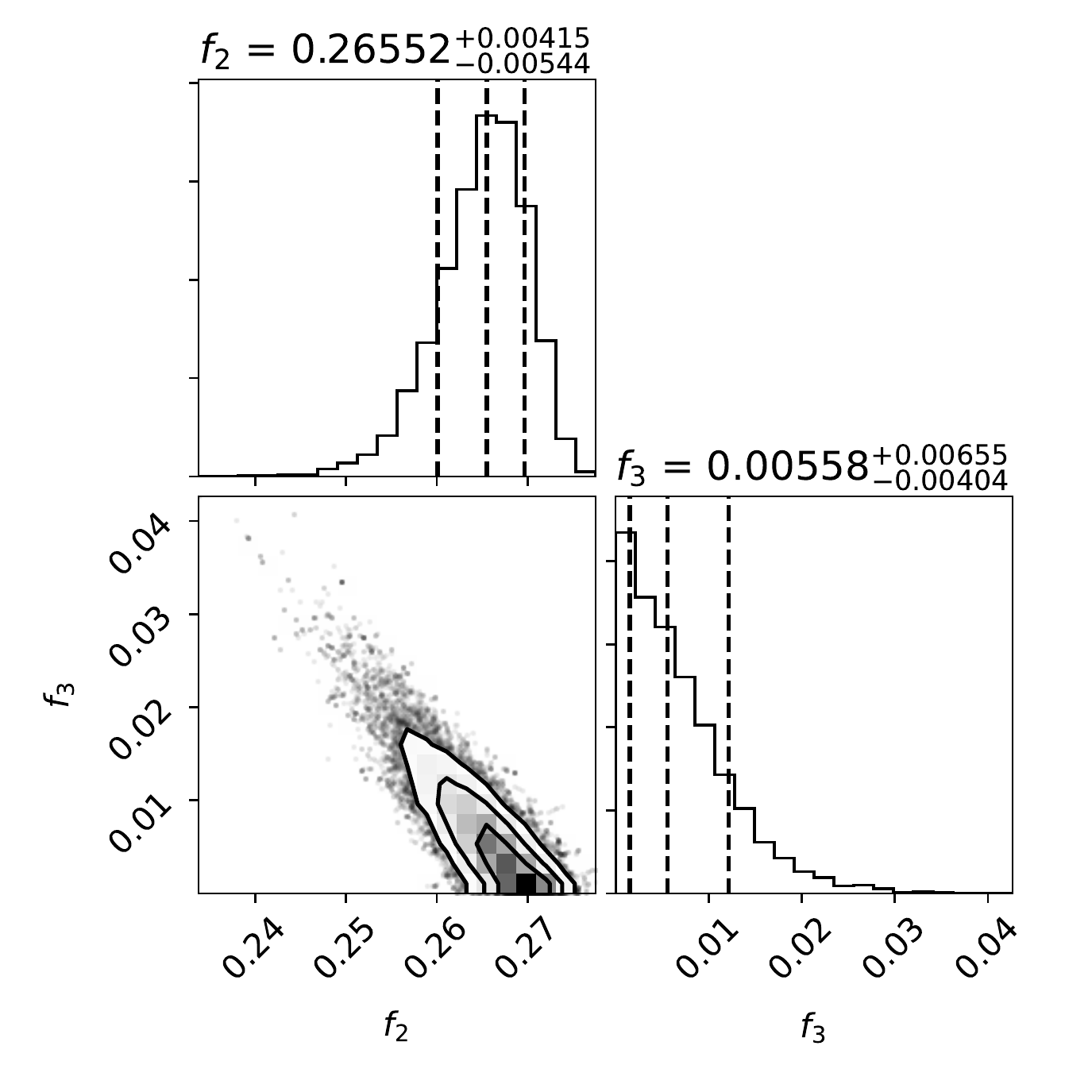}
\includegraphics[width=0.5\textwidth]{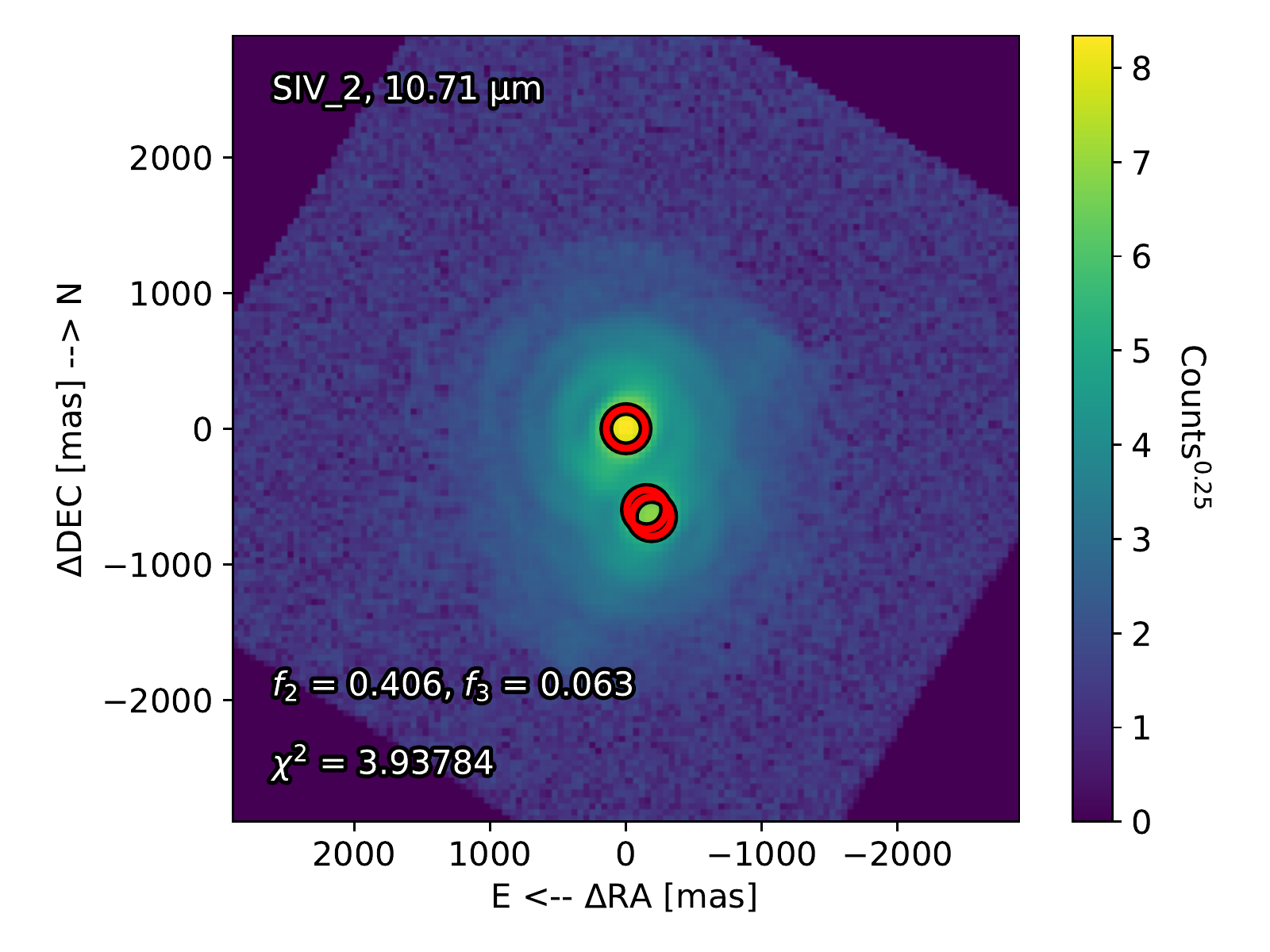}
\includegraphics[width=0.4\textwidth]{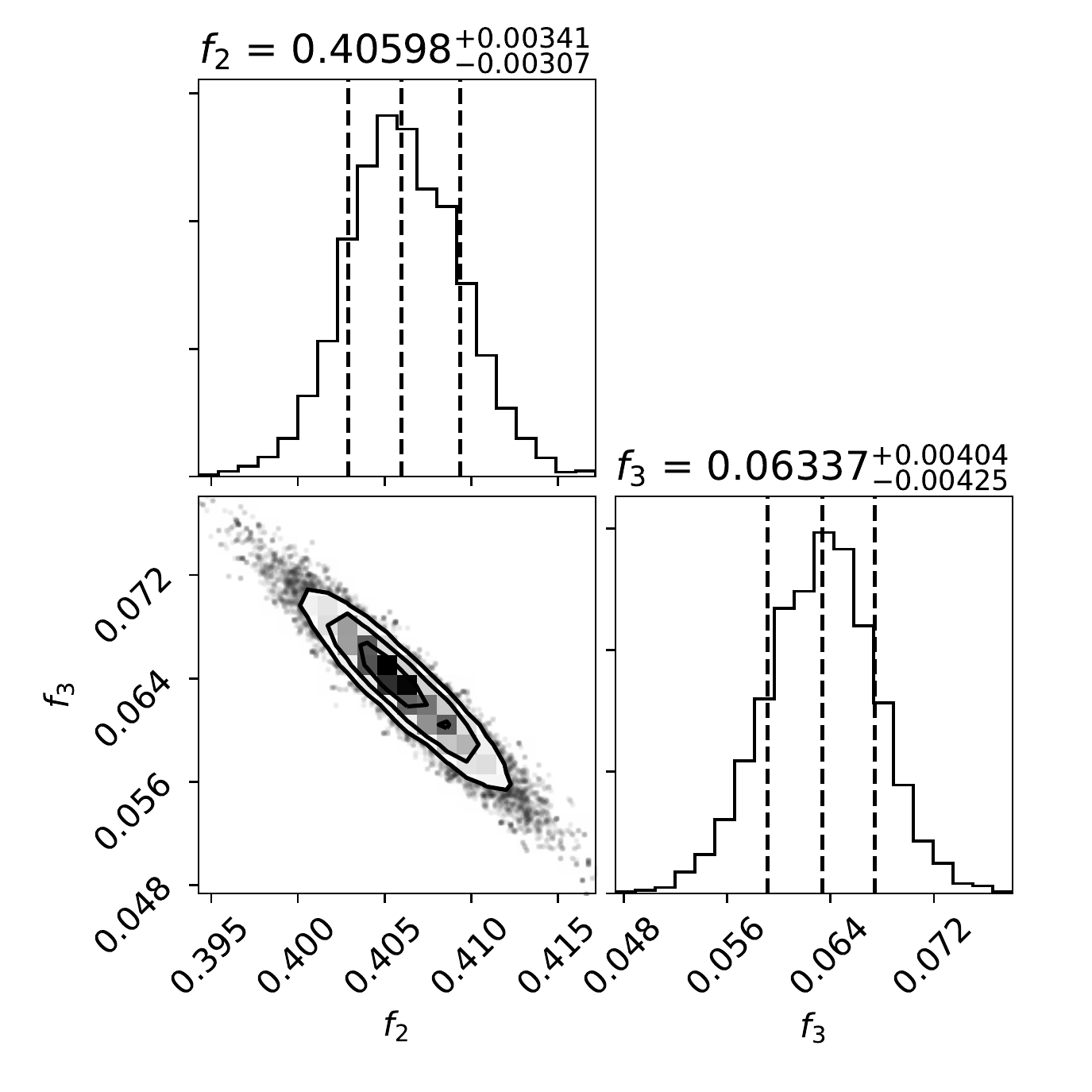}
\caption{De-rotated and co-added VISIR-NEAR images of the T Tau triple system and the three-component fit (left-hand column, red circles) and the corresponding MCMC corner plots \citep[][right-hand column]{foreman-mackey2016} in the ARIII, SIV\_1, and SIV\_2 filters (from top to bottom).}
\label{fig:triple_model_fits}
\end{figure*}

Figure~\ref{fig:triple_model_fits} shows the results of our triple model fits to the T Tau system, with only the fluxes of T Tau Sa and Sb relative to T Tau N left as free parameters. We did both a simple minimization based on a Broyden–Fletcher–Goldfarb–Shanno least-squares algorithm (left-hand panels) and an MCMC fit using \texttt{emcee}\footnote{\url{https://github.com/dfm/emcee}} \citep[][right-hand panels]{foreman-mackey2013}. In both cases, we used half of the combined relative flux of T Tau S as initial values for the fluxes of T Tau Sa and Sb relative to T Tau N. Moreover, we restricted the search space to $0 \leq f_{2,3} \leq 1$ and normalized the log-likelihood function minimized with the MCMC algorithm by a temperature equal to the reduced $\chi^2$ of the best fit triple model (see the left-hand panels of Figure~\ref{fig:triple_model_fits}) in order to artificially inflate the uncertainties. As can be seen in the corner plots, there is a strong anti-correlation between the relative fluxes of T Tau Sa and Sb caused by the fact that their combined flux is constrained by the data. Furthermore, T Tau Sb is fainter than T Tau Sa in all three filters, which is similar to the results from \citet{skemer2008}, and it is consistent with providing zero flux at $9.81~\text{\textmu m}$.

\begin{figure*}
\centering
\includegraphics[width=\textwidth]{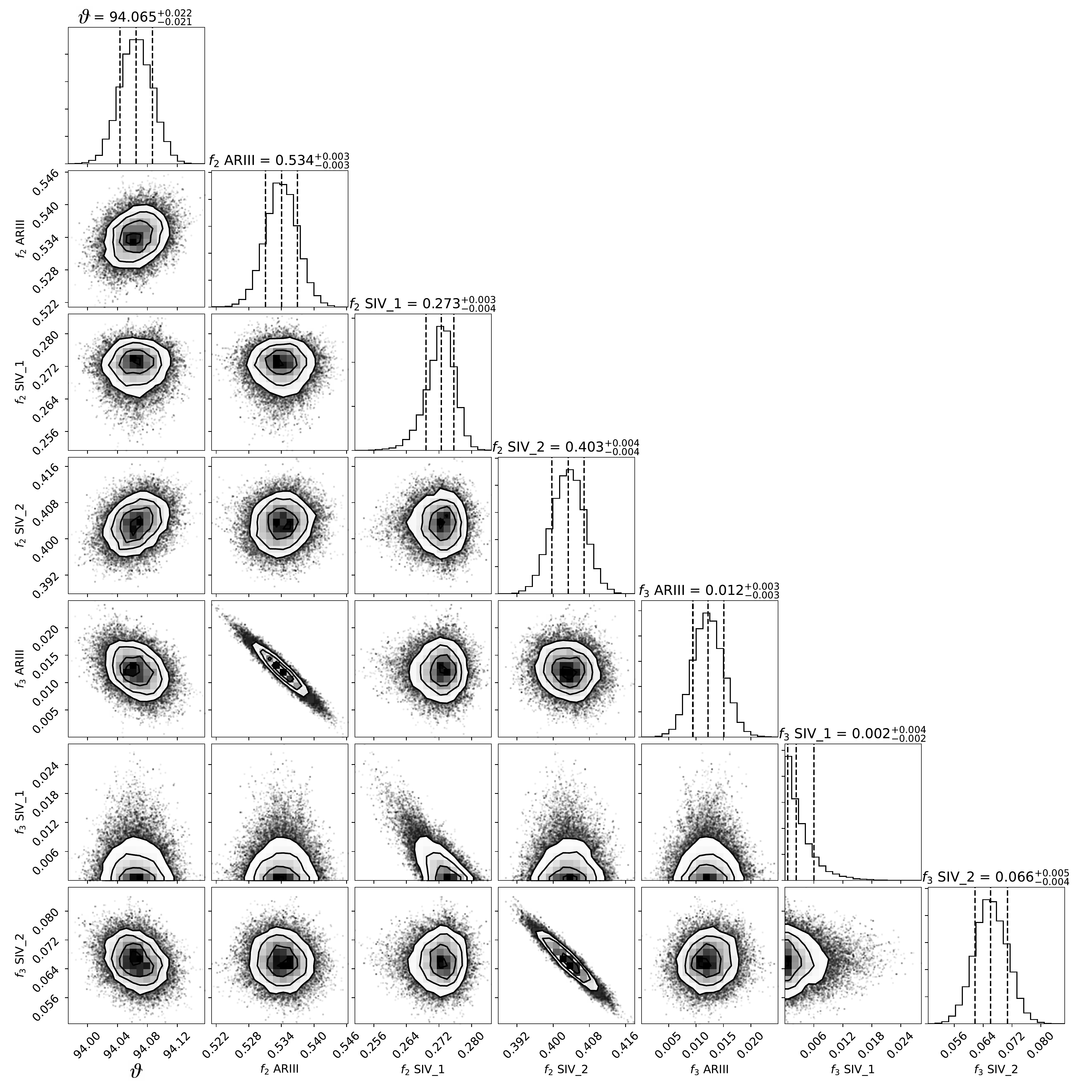}
\caption{Corner plot \citep{foreman-mackey2016} of an MCMC fit of the kernel phase triple model with the rotation offset $\phi$ and the relative fluxes of T Tau Sa ($f_2$) and Sb ($f_3$) as free parameters, simultaneously for the ARIII, SIV\_1, and SIV\_2 data.}
\label{fig:all_fit}
\end{figure*}

Since we estimated the accuracy of the fixed VISIR-NEAR position angle offset $\vartheta$ determined from $\alpha$ Centauri data to be $\sim 0.10~\text{deg}$, we repeated the MCMC fit with $\vartheta$ and the fluxes of T Tau Sa and Sb relative to T Tau N as free parameters, simultaneously for the ARIII, SIV\_1, and SIV\_2 data (see Figure~\ref{fig:all_fit}). This approach allows us to explore the entire parameter space and to assess the errors originating from the uncertainty in the position angle offset $\vartheta$. The results are similar to when the position angle offset $\vartheta$ is fixed and each filter is fitted separately, with similar uncertainties, suggesting that the uncertainty in the position angle offset $\vartheta$ does not significantly limit our precision. We note that the best fit position angle offset $\vartheta_\text{fit} = 94.07 \pm 0.02$ is consistent with our assumption of $\vartheta_\text{prior} = 94.02 \pm 0.10$ that is based on $\alpha$ Centauri data.

Finally, we repeated the simultaneous triple model fits while varying the position offsets of T Tau Sa and Sb by $\pm 1~\text{mas}$ in order to empirically estimate the errors originating from the $\sim 1~\text{mas}$ uncertainty in these offsets. Again, we find similar results within the uncertainties, suggesting that the uncertainty in the position offsets of T Tau Sa and Sb does not significantly limit our precision  either.

\subsection{NEII data}
\label{sec:neii_data}

In the NEII data, at $12.80~\text{\textmu m}$, the southern binary of the T Tau triple system is separated by only $0.2~\lambda/D$. Moreover, it is brighter than T Tau N, so we had to center the frames on T Tau S before extracting the kernel phase (only if the brightest PSF is in the center of the frames can we avoid the Fourier phase $\phi$ wrapping around $\pm \pi$ and leading to discontinuities and a corrupt kernel phase $\theta$).

In order to perfectly center the frames on T Tau Sa, we first used \texttt{XARA}'s sub-pixel re-centering routine to center the frames on T Tau N. We then computed the exact shift that we have to apply to the frames in order to put T Tau Sa into the center. This is possible because we know its $(\Delta_\text{RA},\Delta_\text{DEC})$ offset from VLT/SPHERE astrometry. Directly centering on T Tau Sa is impossible since its PSF is unresolved with the PSF of T Tau Sb; as such, \texttt{XARA}'s re-centering routine would only find the combined photometric center of T Tau Sa and Sb. Then, we proceeded similarly to how we treated the ARIII, SIV\_1, and SIV\_2 data and fit the kernel phase binary and triple models to the data. However, for the triple model fits, the offsets of T Tau N and Sb relative to the center of the frames (i.e., T Tau Sa) are now (648.1~mas N, 188.5~mas E) and (52.5~mas N, 37.1~mas E).

\begin{figure*}
\centering
\includegraphics[width=0.5\textwidth]{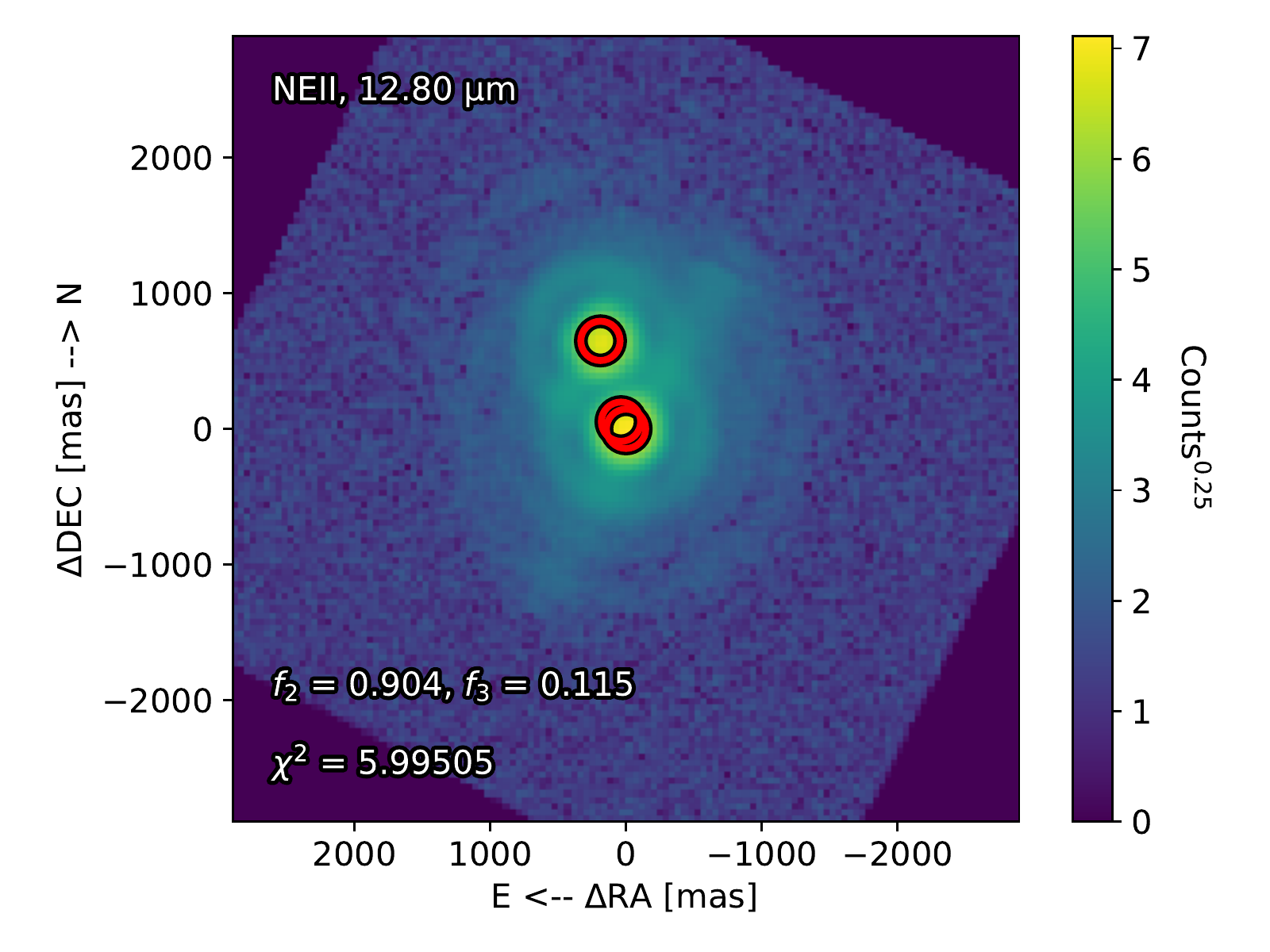}
\includegraphics[width=0.4\textwidth]{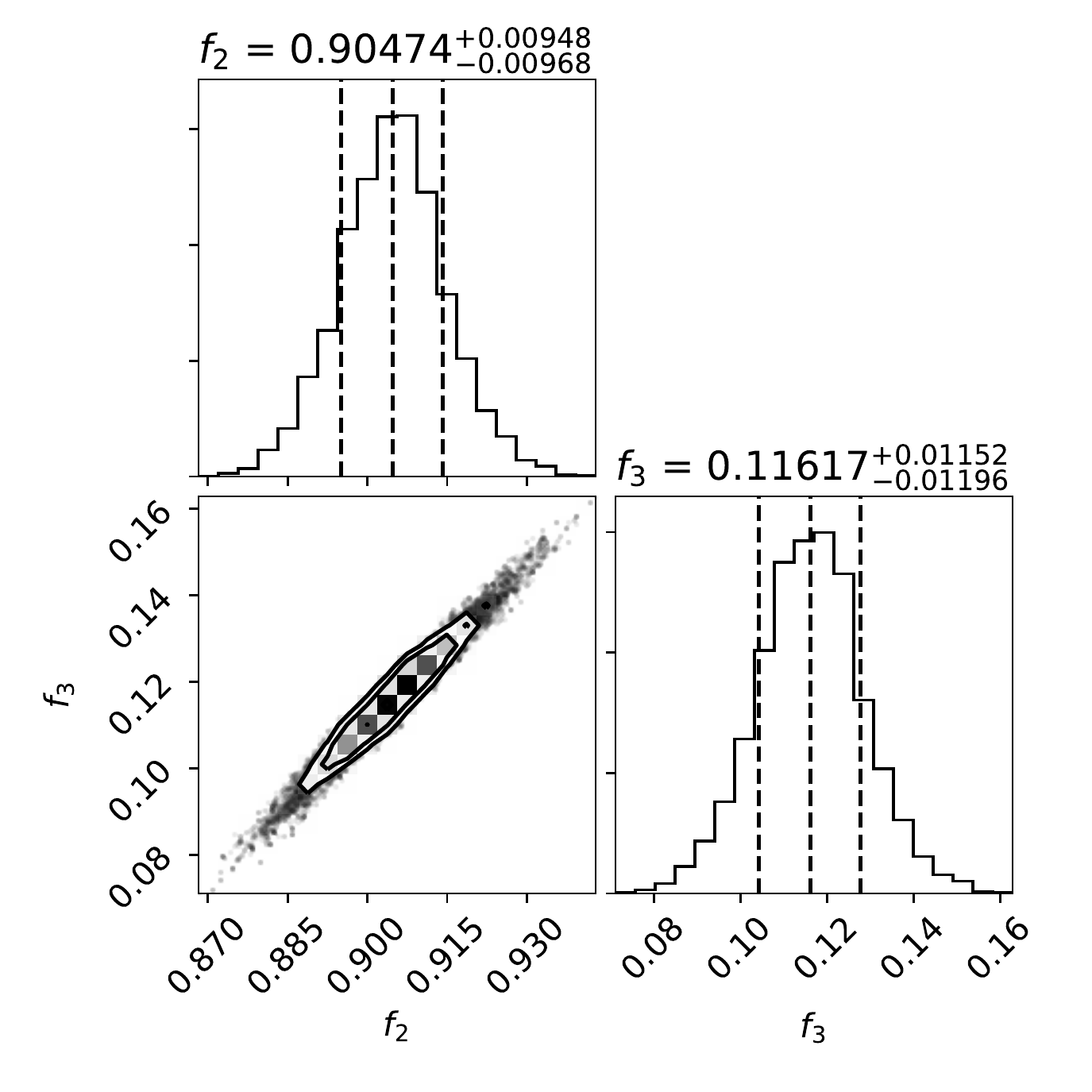}
\caption{Same as Figure~\ref{fig:triple_model_fits} but for the NEII filter and with the relative fluxes of T Tauri N ($f_2$) and Sb ($f_3$) as free parameters.}
\label{fig:neii_data}
\end{figure*}

Our triple model fit results are shown in Figure~\ref{fig:neii_data}. We note that there is now a positive correlation between the relative fluxes of T Tau N and Sb since they are both measured relative to the same PSF of T Tau Sa and are well resolved from each other.

\subsection{Mock data}
\label{sec:mock_data}

In order to validate our approach, we created mock data of the T Tau system and tried to recover it using our kernel phase model fitting techniques. Therefore, we used the calibrator HD~27639 as a reference PSF. Then, we shifted, normalized, and co-added this reference PSF so that it models the three components of the T Tau system. For each filter, we modeled one system with the observed flux ratios and three more systems with T Tau Sa/Sb flux ratios of 0.5/0.5, 0.75/0.25, and 0.9/0.1 (T Tau Sa/Sb flux ratios of 1/0.5, 1/0.25, and 1/0.1 for the NEII filter). In all scenarios, the combined flux of T Tau Sa and Sb relative to T Tau N (the flux of T Tau N for the NEII filter) is equal to the value that we obtained from the binary model fit (see Figure~\ref{fig:binary_model_fits}). We applied the kernel phase technique to the mock data in exactly the same fashion  as we did to the real data (e.g., a different re-centering method for the ARIII, SIV\_1, and SIV\_2 data compared to the NEII data).

Table~\ref{tab:sim_fit} reports the simulated and recovered relative fluxes for all scenarios and the ARIII, SIV\_1, SIV\_2, and NEII filters. The initial values for $f_2$ and $f_3$ were always set to half of the combined flux of T Tau S relative to T Tau N. We note that for the NEII filter, the relative flux of T Tau Sa is fixed to one and we are fitting for the relative fluxes of T Tau N and Sb.

In most cases, the recovered relative fluxes are consistent with the simulated ones within two sigma, and in all cases within three sigma. Given that the fits with mock data do not account for systematic errors and that our covariance model underestimates the uncertainties (see Section~\ref{sec:binary_model_fits}), differences of two to three sigma are expected. This leads us to the conclusion that the photometry obtained from our kernel phase model fitting techniques is reliable down to angular separations of $\sim 0.2~\lambda/D$ (that is the angular separation of T Tau Sa and Sb at the longest wavelength in the NEII filter). However, it is also clear that our techniques are limited by systematic errors, and not the uncertainty in the fixed position angle offset $\vartheta$ or the $(\Delta_\text{RA},\Delta_\text{DEC})$ offset of T Tau Sa and Sb relative to T Tau N obtained from VLT/SPHERE astrometry. This can be concluded from the relatively large discrepancies between the simulated and recovered fluxes when compared to the uncertainties predicted by the MCMC fit, which accounts for the largest statistical noise (photon noise). Hence, we empirically estimated the uncertainties of the relative fluxes from the observed differences between the simulated and recovered values for the mock data (see Table~\ref{tab:triple_model_fits}). We adapted a conservative uncertainty of $\sigma_\text{emp} = 0.02$ for the ARIII, SIV\_1, and SIV\_2 filters, given that all observed differences fall within this range, and a slightly larger uncertainty of $\sigma_\text{emp} = 0.03$ for the NEII filter, given the slightly larger observed differences due to the longer wavelength and the different data reduction approach.

\subsection{Photometry}
\label{sec:photometry}

\begin{figure*}
\centering
\includegraphics[width=\textwidth]{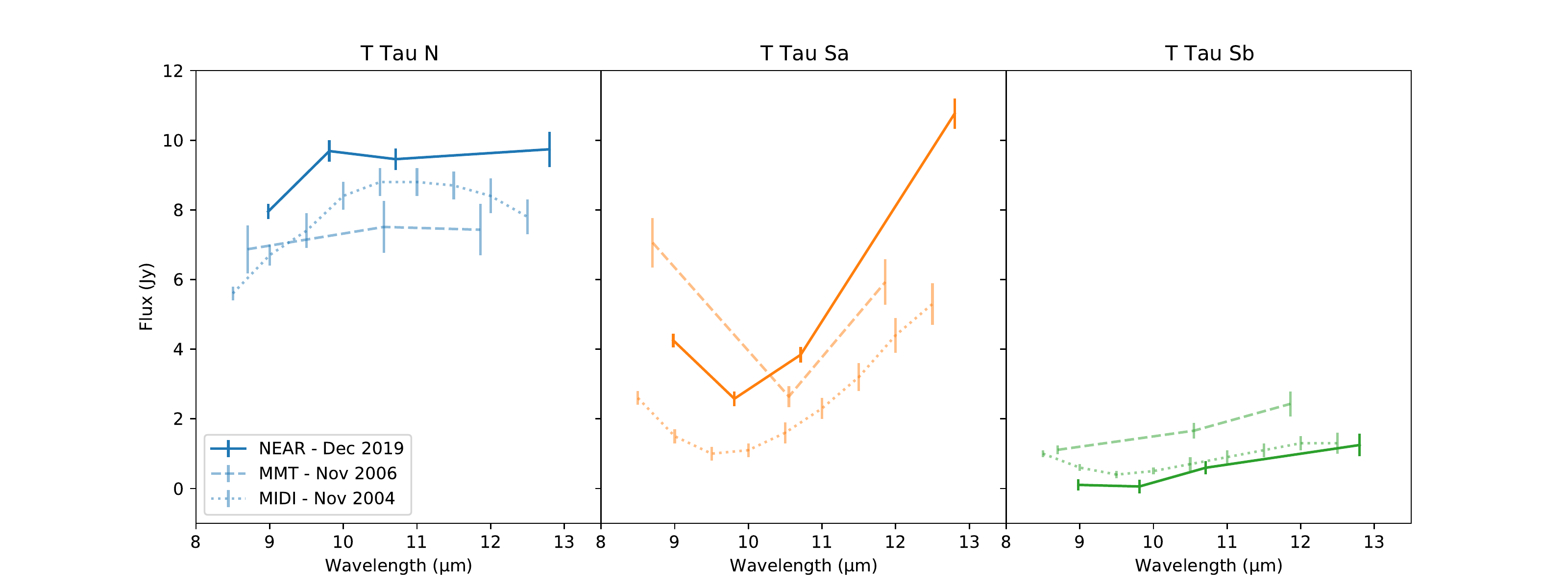}
\caption{Photometry of T Tau N (blue curves), Sa (orange curves), and Sb (green curves) from $\sim 8$--$13~\text{\textmu m}$. There are VISIR-NEAR data from December 2019 (solid curves, this work), MMT data from November 2006 \citep[dashed curves,][]{skemer2008}, and MIDI data from November 2004 \citep[dotted curves,][]{ratzka2009}.}
\label{fig:fluxes}
\end{figure*}

The final photometry of T Tau N, Sa, and Sb was computed from the kernel phase relative fluxes and the PSF photometry with an aperture of 70 pixels in diameter on the cleaned frames of T Tau and HD~27639 (the calibrator). HD~27639 is part of the VISIR\footnote{\url{http://www.eso.org/sci/facilities/paranal/instruments/visir/tools/zerop_cohen_Jy.txt}} photometric standards list, from which we adapted fluxes of $14.9 \pm 0.3~\text{Jy}$ (ARIII), $13.1 \pm 0.3~\text{Jy}$ (SIV\_1), $11.3 \pm 0.3~\text{Jy}$ (SIV\_2), and $8.7 \pm 0.3~\text{Jy}$ (NEII) with conservative uncertainties. Our final photometry for the T Tau system, using the results from Figures~\ref{fig:triple_model_fits} and~\ref{fig:neii_data} for the relative fluxes, is reported in Table~\ref{tab:triple_model_fits} and shown in Figure~\ref{fig:fluxes}.

\begin{table}
\caption{Best fit relative fluxes of the T Tau system from an MCMC fit of the kernel phase triple model to the VISIR-NEAR data. We also report an empirical uncertainty ($\sigma_\text{emp}$) for the relative fluxes (see Section~\ref{sec:mock_data}) as well as the final photometry of T Tau N, Sa, and Sb in all filters.}
\label{tab:triple_model_fits}
\centering
\begin{tabular}{c r c l l r}
\hline\hline
Filter & $\lambda$ [$\text{\textmu m}$] & Star & Rel. flux & $\sigma_\text{emp}$ & Flux [Jy] \\
\hline
ARIII & 8.98 & N & 1.000 & 0 & $8.12 \pm 0.13$ \\
ARIII & 8.98 & Sa & $0.535^{+0.003}_{-0.003}$ & 0.02 & $4.34 \pm 0.11$ \\
ARIII & 8.98 & Sb & $0.013^{+0.003}_{-0.003}$ & 0.02 & $0.10 \pm 0.08$ \\
\hline
SIV\_1 & 9.81 & N & 1.000 & 0 & $9.81 \pm 0.16$ \\
SIV\_1 & 9.81 & Sa & $0.266^{+0.004}_{-0.005}$ & 0.02 & $2.68 \pm 0.07$ \\
SIV\_1 & 9.81 & Sb & $0.006^{+0.007}_{-0.004}$ & 0.02 & $0.02 \pm 0.05$ \\
\hline
SIV\_2 & 10.71 & N & 1.000 & 0 & $10.04 \pm 0.19$ \\
SIV\_2 & 10.71 & Sa & $0.406^{+0.003}_{-0.003}$ & 0.02 & $4.05 \pm 0.13$ \\
SIV\_2 & 10.71 & Sb & $0.063^{+0.004}_{-0.004}$ & 0.02 & $0.66 \pm 0.10$ \\
\hline
NEII & 12.80 & N & $0.905^{+0.009}_{-0.010}$ & 0.03 & $9.63 \pm 0.43$ \\
NEII & 12.80 & Sa & 1.000 & 0 & $10.64 \pm 0.42$ \\
NEII & 12.80 & Sb & $0.116^{+0.012}_{-0.012}$ & 0.03 & $1.23 \pm 0.32$ \\
\hline
\end{tabular}
\end{table}

\section{Discussion}
\label{sec:discussion}

Figure~\ref{fig:fluxes} shows the photometry of T Tau N, Sa, and Sb in the mid-infrared over a period of $\sim$15 years. All three components were resolved with VLTI/MIDI interferometry in November 2004 \citep{ratzka2009}, with the MMT Mid-IR Array Camera 4 (MIRAC4) using high-contrast imaging in November 2006 \citep{skemer2008}, and VLT/VISIR-NEAR kernel phase interferometry (this work) in December 2019.

The SED of T Tau N is quite stable over time, within the uncertainties, and confirms the consistent photometry typically obtained from this star. We also confirm the detection of weak silicate emission \citep{ghez1991}, which suggests the presence of a face-on CSD around T Tau N. Instead, the SED of T Tau Sa varies by up to a factor of four over the N-band, with our measurement falling approximately in between the values reported by \citet{skemer2008} and \citet{ratzka2009} at the shorter wavelengths. However, our $12.8~\text{\textmu m}$ flux falls significantly below the N-band fluxes measured in October 1990 ($17.1 \pm 2.3~\text{Jy}$) and February 2008 ($12.1$--$16.7~\text{Jy}$) by \citet{ghez1991} and \citet{vanboekel2010}, respectively. So, while T Tau Sa is currently rather bright in the near-infrared \citep{schaefer2020} when compared to the 2004--2006 period, its N-band brightness is average at best. This is the well-known "bluer when brighter" behavior typical for extinction, but it is also consistent with variable accretion from a small edge-on CSD around T Tau Sa \citep{vanboekel2010}. Variability in the $\sim 9.7~\text{\textmu m}$ silicate feature has also been observed for other strongly accreting CTTSs, such as CW Tau, DG Tau, and XZ Tau \citep{leisenring2007}.

For T~Tau~Sa, we also detect the silicate absorption feature known to exist in the $10~\text{\textmu m}$ spectral region around the southern component \citep{ghez1991}. This feature was later shown to arise from extinction toward T Tau Sa and Sb, but with a greater optical depth toward Sa \citep{ratzka2009}. As pointed out by these authors, this difference in extinction must have been rather local given the small projected separation between T Tau Sa and Sb of $\sim 0.1~\text{arcsec}$ or $\sim 15~\text{au}$, and they attributed it to silicate absorption caused by the outer parts of T Tau Sa's CSD. However, \citet{vanboekel2010} showed that Sa's tidally truncated small disk should be warm enough even in its outer parts to produce silicate emission, including when seen edge-on. Therefore, additional extinction caused by, for example, the southern CBD is necessary to explain the SED of T Tau Sa.

In contrast to T Tau Sa, our data show a significant dimming of T Tau Sb when compared to the 2004--2006 period. The effect is most prominent around the silicate feature, where Sb has been dimming by at least a factor of five or $\sim 2~\text{mag}$ (see Figure~\ref{fig:fluxes}). This is consistent with the current K-band dimming of $\sim 2$--$2.5~\text{mag}$ observed over the same period \citep{schaefer2020} and enables us to infer an upper limit on the dust size by considering standard dust models \citep[see e.g.,][]{kruegel1994} with a size distribution of $n(a) \sim a^{-3.5}$ \citep{mathis1977}, which predict a bell-shaped silicate feature with a maximum near $9.7~\text{\textmu m}$. For these, the larger the maximum grain size to cut the power-law distribution is set, the wider and shallower the silicate feature becomes. For a maximum grain size well below $1~\text{\textmu m}$, which is typical for interstellar dust not processed in the dense inner regions of CSDs, the dust opacity around $10~\text{\textmu m}$ would be similar to that in the K-band, and therefore the extinction would also be similar.

\citet{koehler2020} proposed that T~Tau~Sb has now moved along its orbit around Sa through the southern CBD and consequently suffers increased dust extinction. The Sa-Sb CBD is aligned roughly north to south, with an inclination of $\sim 60~\text{deg}$ and a position angle of $\sim 30~\text{deg}$ \citep[see Figure~1 of][]{koehler2020}. Therefore, T Tau Sb was located behind the dense inner region of the Sa-Sb CBD, which extents out to $\sim 150~\text{mas}$ \citep{yang2018} from the center of mass ($\sim$ T Tau Sa), at the time of our observations. Our photometry thus serves as evidence for increased dust extinction toward T Tau Sb caused by its orbital motion through the southern CBD and confirms the scenario proposed by \citet{koehler2020}. Another possible explanation would be a cloud of dust entering the observing beam in front of T Tau Sb. Such a clumpy structure would be motivated by the strong outflows and turbulence observed in the T Tau system \citep[e.g.,][]{kasper2016}. However, since T Tau Sb was already observed to be faint by \citet{koresko2000} in data from 1997 \citep[when it was on the opposite side of T Tau Sa; see Figure~1 of][]{koehler2020}, it seems more likely that the Sa-Sb CBD is responsible for its dimming.

Comparing our mid-infrared photometry from December 2019 to that from November 2006 and 2004, we see the silicate absorption variability of T Tau Sb that was introduced by its passage through the southern CBD, but we do not claim variability of T Tau Sa's silicate feature. We note that \citet{vanboekel2010} measured variability at $12.8~\text{\textmu m}$ (i.e., not in the silicate feature) of the unresolved T Tau S binary over just a few days and attributed this to variable accretion, where UV emission is reprocessed by the disk and shows up in the mid-infrared. These authors also modeled T Tau Sa's truncated CSD and found that they cannot reproduce silicate absorption by the CSD alone (instead, Sa's small and warm CSD always shows silicate in emission) and that foreground extinction by cold dust (e.g., by the CBD) is therefore needed.

\section{Conclusions}
\label{sec:conclusions}

In this paper, we measure the brightness of all three components of the iconic T Tau system in the mid-infrared, around $\sim 10~\text{\textmu m}$, in order to perform photometry and study the variability caused by its dynamic and dusty environment. We confirm weak silicate emission around T Tau N, indicating a face-on CSD, and strong silicate absorption around T Tau Sa, consistent with extinction by southern circumbinary material \citep[e.g.,][]{vanboekel2010}. For T Tau Sb, we observe a dimming over the entire $\sim 8$--$13~\text{\textmu m}$ spectral range when compared to data from 2004 and 2006, which is in agreement with recent dimming in the near-infrared (K-band) observed by \citet{schaefer2020}.

The dimming of T Tau Sb observed over a wide spectral range (near- to mid-infrared) is consistent with Sb having moved behind the dense inner region of the Sa-Sb CBD, resulting in increased dust extinction. Therefore, we can confirm the scenario proposed by \citet{koehler2020} and the Sa-Sb CBD geometry derived by \citet{yang2018}, which, together with previous works \citep[e.g.,][]{duchene2002,skemer2008,ratzka2009}, strengthen the evidence for a significant misalignment between the CBD and the T Tau Sa CSD. Most surprisingly, these disks are also misaligned with the orbit of T Tau Sb around Sa and T Tau Sb's non-edge-on disk, regardless of their small physical separation of $\sim 15~\text{au}$. At such small separations, tidal forces should align them within short timescales \citep{ratzka2009}, challenging star-formation models and suggesting that multiple star formation can be turbulent \citep[e.g.,][]{whitworth2001,jensen2004}.

At angular separations down to $\sim 0.2~\lambda/D$, we use kernel phase interferometry \citep{martinache2010} together with the known positions of T Tau N, Sa, and Sb \citep{koehler2016} to robustly determine their brightnesses. We validate our methods by simulating and recovering mock data of the T Tau triple system using the observed PSF reference. We find that kernel phase interferometry, applied to high-Strehl images in the mid-infrared, is a powerful technique to achieve the highest possible angular resolution with single-dish telescopes. The adaptive optics-corrected VISIR-NEAR instrument provides a unique opportunity to perform such observations from the ground, while kernel phase will also find increased application in space-based observations with the \emph{James Webb Space Telescope} \citep[e.g.,][]{ceau2019}.

\begin{acknowledgements}
The NEAR project was made possible through contributions from the Breakthrough Foundation and Breakthrough Watch program, as well as through contributions from the European Southern Observatory. The manuscript was also substantially improved following helpful comments from an anonymous referee.
\end{acknowledgements}

%
%

\bibliographystyle{aa}
\bibliography{references}

\begin{appendix}

\section{Tests with mock data}

\begin{table*}
\caption{Simulated and recovered relative fluxes for our mock data of the T Tau triple system. The values of $f_2$ and $f_3$ correspond to T Tau Sa and Sb for the ARIII, SIV\_1, and SIV\_2 filters and to T Tau N and Sb for the NEII filter, respectively. We perform fits using both a Broyden–Fletcher–Goldfarb–Shanno least-squares algorithm (minimize) and an MCMC algorithm (\texttt{emcee}).}
\label{tab:sim_fit}
\centering
\begin{tabular}{c c r r r r r r}
\hline\hline
Filter & Test no. & Sim. $f_2$ & Rec. $f_2$ (minimize) & Rec. $f_2$ (\texttt{emcee}) & Sim. $f_3$ & Rec. $f_3$ (minimize) & Rec. $f_3$ (\texttt{emcee}) \\
\hline
\multirow{4}{*}{ARIII} & 1 & 0.534 & 0.530 & $0.530^{+0.003}_{-0.003}$ & 0.012 & 0.009 & $0.010^{+0.002}_{-0.002}$ \\
& 2 & 0.273 & 0.271 & $0.271^{+0.003}_{-0.002}$ & 0.273 & 0.269 & $0.269^{+0.002}_{-0.002}$ \\
& 3 & 0.410 & 0.407 & $0.407^{+0.003}_{-0.003}$ & 0.137 & 0.133 & $0.133^{+0.002}_{-0.002}$ \\
& 4 & 0.491 & 0.488 & $0.488^{+0.002}_{-0.003}$ & 0.055 & 0.052 & $0.052^{+0.002}_{-0.002}$ \\
\hline
\multirow{4}{*}{SIV\_1} & 1 & 0.273 & 0.267 & $0.254^{+0.011}_{-0.014}$ & 0.002 & 0.000 & $0.010^{+0.011}_{-0.007}$ \\
& 2 & 0.135 & 0.143 & $0.142^{+0.021}_{-0.021}$ & 0.135 & 0.121 & $0.122^{+0.017}_{-0.018}$ \\
& 3 & 0.202 & 0.208 & $0.208^{+0.022}_{-0.023}$ & 0.068 & 0.056 & $0.056^{+0.018}_{-0.018}$ \\
& 4 & 0.242 & 0.247 & $0.240^{+0.018}_{-0.020}$ & 0.027 & 0.017 & $0.023^{+0.016}_{-0.014}$ \\
\hline
\multirow{4}{*}{SIV\_2} & 1 & 0.403 & 0.401 & $0.401^{+0.005}_{-0.005}$ & 0.066 & 0.064 & $0.064^{+0.005}_{-0.006}$ \\
& 2 & 0.232 & 0.232 & $0.232^{+0.005}_{-0.005}$ & 0.232 & 0.229 & $0.229^{+0.005}_{-0.005}$ \\
& 3 & 0.348 & 0.347 & $0.347^{+0.005}_{-0.005}$ & 0.116 & 0.113 & $0.114^{+0.005}_{-0.006}$ \\
& 4 & 0.418 & 0.416 & $0.416^{+0.005}_{-0.005}$ & 0.046 & 0.044 & $0.044^{+0.006}_{-0.005}$ \\
\hline
\multirow{4}{*}{NEII} & 1 & 0.905 & 0.918 & $0.919^{+0.009}_{-0.009}$ & 0.116 & 0.140 & $0.141^{+0.012}_{-0.011}$ \\
& 2 & 0.905 & 0.920 & $0.920^{+0.019}_{-0.017}$ & 0.500 & 0.535 & $0.535^{+0.028}_{-0.027}$ \\
& 3 & 0.905 & 0.920 & $0.921^{+0.012}_{-0.011}$ & 0.250 & 0.279 & $0.281^{+0.015}_{-0.015}$ \\
& 4 & 0.905 & 0.918 & $0.918^{+0.009}_{-0.008}$ & 0.100 & 0.124 & $0.124^{+0.011}_{-0.011}$ \\
\hline
\end{tabular}
\end{table*}

\end{appendix}

\end{document}